\documentclass[useAMS,usenatbib]{mn2e}

\usepackage{rotating}
\usepackage{longtable}
\usepackage{graphics}
\usepackage{aas_macros}

\newcommand{\teff}{$T_{\rm eff}$}
\newcommand{\dnlte}{$\rm \Delta_{NLTE}$}
\newcommand{\eexc}{$E_{\rm exc}$}
\def\lgg{log\,${g}$}
\def\vt{$\xi_{\rm t}$}

\newcommand{\kms}{km\,s$^{-1}$}

\def\ione{\,{\sc i}}
\def\ii{\,{\sc ii}}
\def\iii{\,{\sc iii}}
\newcommand{\kH}{$S_{\rm H}$}    

\title[A NLTE line formation for Ti\ione\ and Ti\ii\ in model atmospheres of the reference A-K stars]{A NLTE line formation for neutral and singly-ionised  titanium in model atmospheres of the reference A-K stars}
\author[T. M. Sitnova, L. I. Mashonkina, T. A. Ryabchikova]{T. M. Sitnova$^{1,2}$\thanks{E-mail:
sitnova@inasan.ru}; L. I. Mashonkina$^{1}$, T. A. Ryabchikova$^{1}$ \\
$^{1}$Institute of Astronomy, Russian Academy of Sciences, Pyatnitskaya 48, 119017, Moscow, Russia\\
$^{2}$Sternberg Astronomical Institute; Faculty of Physics, Moscow State University, Universitetsky pr., 13, 119991, Moscow, Russia\\}
\begin{document}

\date{}

\pagerange{\pageref{firstpage}--\pageref{lastpage}} \pubyear{2016}

\maketitle

\label{firstpage}

\begin{abstract}

We construct a model atom for Ti\ione--\ii\ using more than 3600 measured and predicted energy levels of Ti\ione\ and 1800 energy levels of Ti\ii, and quantum mechanical photoionisation cross-sections. Non-local thermodynamical equilibrium (NLTE) line formation for Ti\ione\ and Ti\ii\ is treated through a wide range of spectral types from A to K, including metal-poor stars with [Fe/H] down to $-2.6$~dex. NLTE leads to weakened Ti\ione\ lines and positive abundance corrections. The magnitude of NLTE corrections is smaller compared to the literature data for FGK atmospheres. NLTE leads to strengthened Ti\ii\ lines and negative NLTE abundance corrections. For the first time, we performed the NLTE calculations for Ti\ione--\ii\ in the 6500~K$\leq$~\teff~$\leq$~13000~K range. For four A type stars we derived in LTE an abundance discrepancy of up to 0.22~dex was obtained between Ti\ione\ and Ti\ii\, and it vanishes in NLTE. For other four A-B stars, with only Ti\ii\ lines observed, NLTE leads to decrease of line-to-line scatter. An efficiency of inelastic Ti\ione~+~H\ione\ collisions was estimated from analysis of Ti\ione\ and Ti\ii\ lines in 17 cool stars with $-2.6 \leq$~[Fe/H]~$\leq$~0.0. Consistent NLTE abundances from Ti\ione\ and Ti\ii\ were obtained applying classical Drawinian rates for the stars with \lgg~$\ge$~4.1, and neglecting inelastic collisions with H\ione\ for the VMP giant HD~122563. For the VMP turn-off stars ([Fe/H]~$\leq -2$ and \lgg~$\leq$~4.1), we obtained the positive abundance difference Ti\ione--\ii\ already in LTE and it increases in NLTE. The accurate collisional data for Ti\ione\ and Ti\ii\ are desired to find a clue to this problem.

\end{abstract}


\section{Introduction}

Titanium is observed in lines of two ionisation stages, Ti\ione\ and Ti\ii, in a wide range of spectral types from A to K.
Experimental oscillator strengths ($f_{ij}$) for Ti\ione\ and Ti\ii\ were measured using a common method \citep[][ respectively]{Lawler2013_ti1,Wood2013_ti2}, which permits to use Ti\ione\ and Ti\ii\ lines for determination of accurate titanium abundances and stellar atmosphere parameters. 
\citet{Bergemann2011}  and Bergemann et al. (2012) investigated the non-local thermodynamic equilibrium (NLTE) line-formation for Ti\ione-\ii\ in the atmospheres of cool stars. The first paper presents the NLTE calculations for the Sun and four metal-poor stars with \teff~$\le$~6350~K  while the second one for red supergiants with 3400~K~$\le$~\teff~$\le$~4400~K, $-0.5 \le$~\lgg~$\le$~1.0, and $-0.5 \le$~[Fe/H]~$\le$~0.5. 
\citet{Bergemann2011} found that the deviations from LTE are small in the solar atmosphere, with the abundance difference between NLTE and LTE (the NLTE abundance correction, $\Delta_{\rm NLTE}$) not exceeding 0.11~dex for Ti\ione\ lines.
For the Sun  \citet{Bergemann2011} derived consistent within 0.04~dex NLTE abundances from Ti\ione\ and Ti\ii\ lines.
However, she failed to achieve the Ti\ione /Ti\ii\ ionisation equilibrium for cool metal-poor (MP, $-2.5 \leq$~[Fe/H]~$\leq -1.3$) dwarfs with well-determined atmospheric parameters. \citet{Bergemann2011} suggested that this can be caused by: (i) neglecting high-excitation levels of Ti\ione\ in the used model atom; (ii) using hydrogenic photoionisation cross-sections; (iii) using a rough theoretical approximation \citep{Drawin1968,Drawin1969} for inelastic collisions with hydrogen atoms. 
We eliminate the first two points in this study. 
We still rely on the Drawinian approximation because  accurate laboratory measurements or quantum mechanical calculations  for  inelastic   Ti\ione~$+$~H\ione\ collisions are not available. Poorly-known collisions with H\ione\ atoms is the main source of the  uncertainties in the NLTE results for stars with \teff~$\le$~7000~K.

For the atmospheres hotter than \teff~$\ge$~6500~K the NLTE calculations for Ti\ione--\ii\ were not yet performed, although the observations indicate a discrepancy in LTE abundances between Ti\ione\ and Ti\ii.
For example, \citet{Bikmaev2002} derived under the LTE assumption the abundance difference Ti\ione--Ti\ii\footnote{Here, log~A(X)=log($N_X/N_{tot}$), where N$_{tot}$ is a total number density; X\ione--X\ii\ means difference in abundance derived from lines of X\ione\ and X\ii, log~A(X\ione)--log~A(X\ii).}~=~$-0.17$~dex and $-0.20$~dex for the A-type stars HD~32115 and HD~37954, respectively. 
\citet{Becker1998} performed the NLTE calculations for Ti\ii\  in A-type stars (Vega, supergiants $\eta$~Leo and 41-3712 from M31) and found that NLTE leads to weakened Ti\ii\ lines, with the NLTE abundance corrections being larger for weak lines compared with those calculated for strong lines. 
Using model atom from \citet{Becker1998}, \citet{Przybilla2006} and \citet{Schiller2008} derived the NLTE abundances from lines of Ti\ii\ in BA-type supergiants and concluded that proper NLTE calculations reduce the line-to-line scatter.

We aim to construct a comprehensive model atom of Ti\ione--\ii\ and to treat a reliable method of abundance determination from different lines of  Ti\ione\ and Ti\ii\ in a wide range of stellar spectral types from late B to K, including metal-poor stars. 
First, we test the new model atom employing the stars with \teff~$\ge$~7100~K, where inelastic collisions with hydrogen atoms do not affect the statistical equilibrium (SE). Then, we empirically constrain an efficiency of collisions with H\ione\ from analysis of Ti\ione\ and Ti\ii\ lines in spectra of cool metal-poor  stars.
In total, we analyse titanium lines in  25 well-studied stars.

We present the constructed  model atom and the NLTE mechanism for Ti\ione\ and Ti\ii\  in Section~\ref{method}. Section~\ref{obspar} describes observations and stellar parameters of our stellar sample. The obtained results for hot and cool stars are  considered in Sections~\ref{hots} and \ref{cools}, respectively.
Our conclusions and recommendations are given in Sect.~\ref{con}.

\section{Method of NLTE calculations for Ti\ione--\ii }
\label{method}

In this section we describe the model atom of titanium, the programs used for computing the level populations and spectral line profiles, and mechanisms of departures from LTE for Ti\ione\ and Ti\ii.

\subsection{The model atom} 

\begin{figure*}
	\centering
	\includegraphics[width=\textwidth]{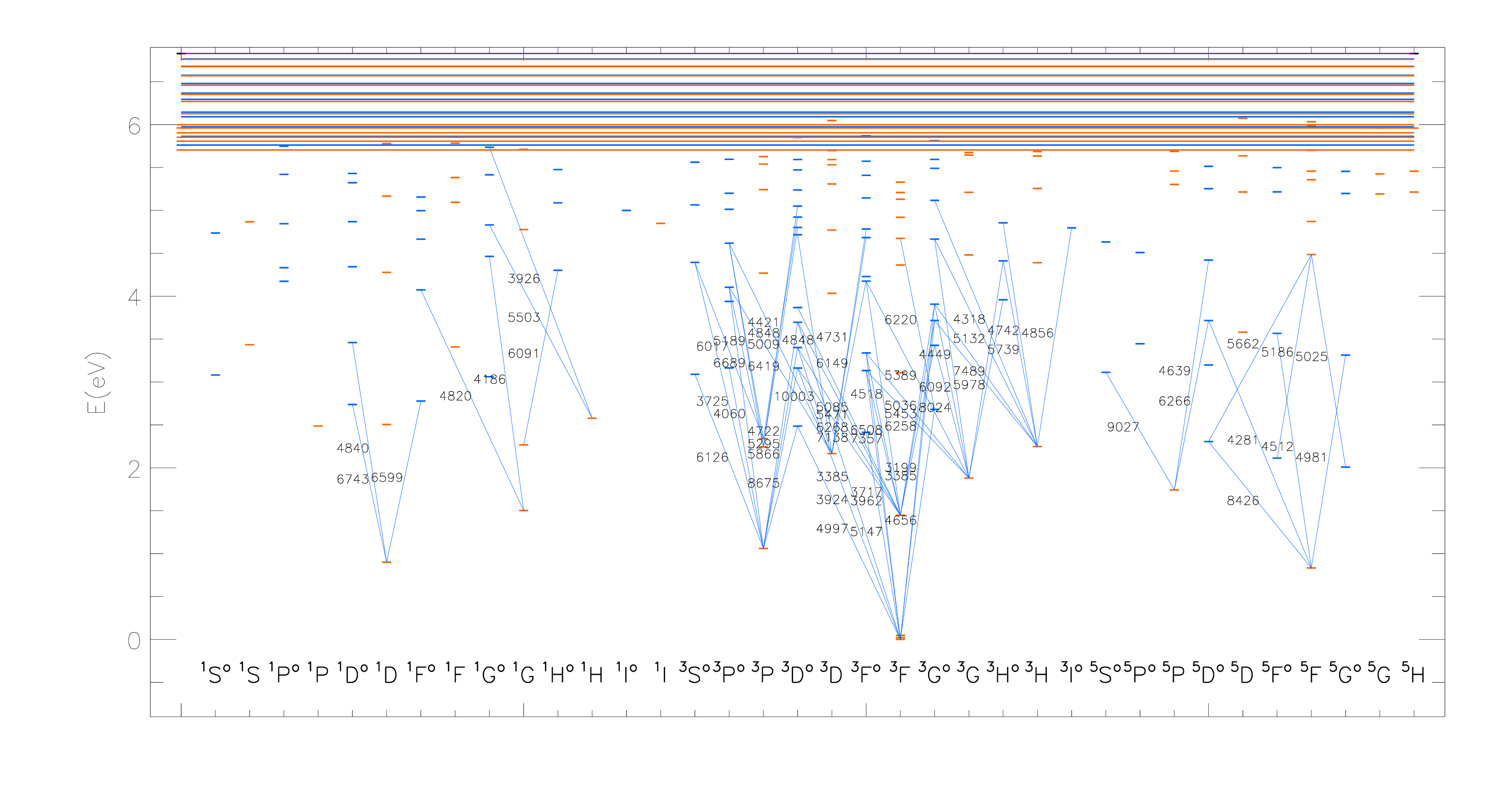}
	\caption{Atomic term structure of Ti\ione\ as obtained from the laboratory experiments (dashes) and calculations (lines). See text for the sources of data. The spectral lines used in abundance analysis arise in the transitions shown as continuous lines. The ionisation threshold for Ti\ione~is~6.83~eV.} 
	\label{ti1} 
\end{figure*}

\begin{figure*}
	\centering
	\includegraphics[width=\textwidth]{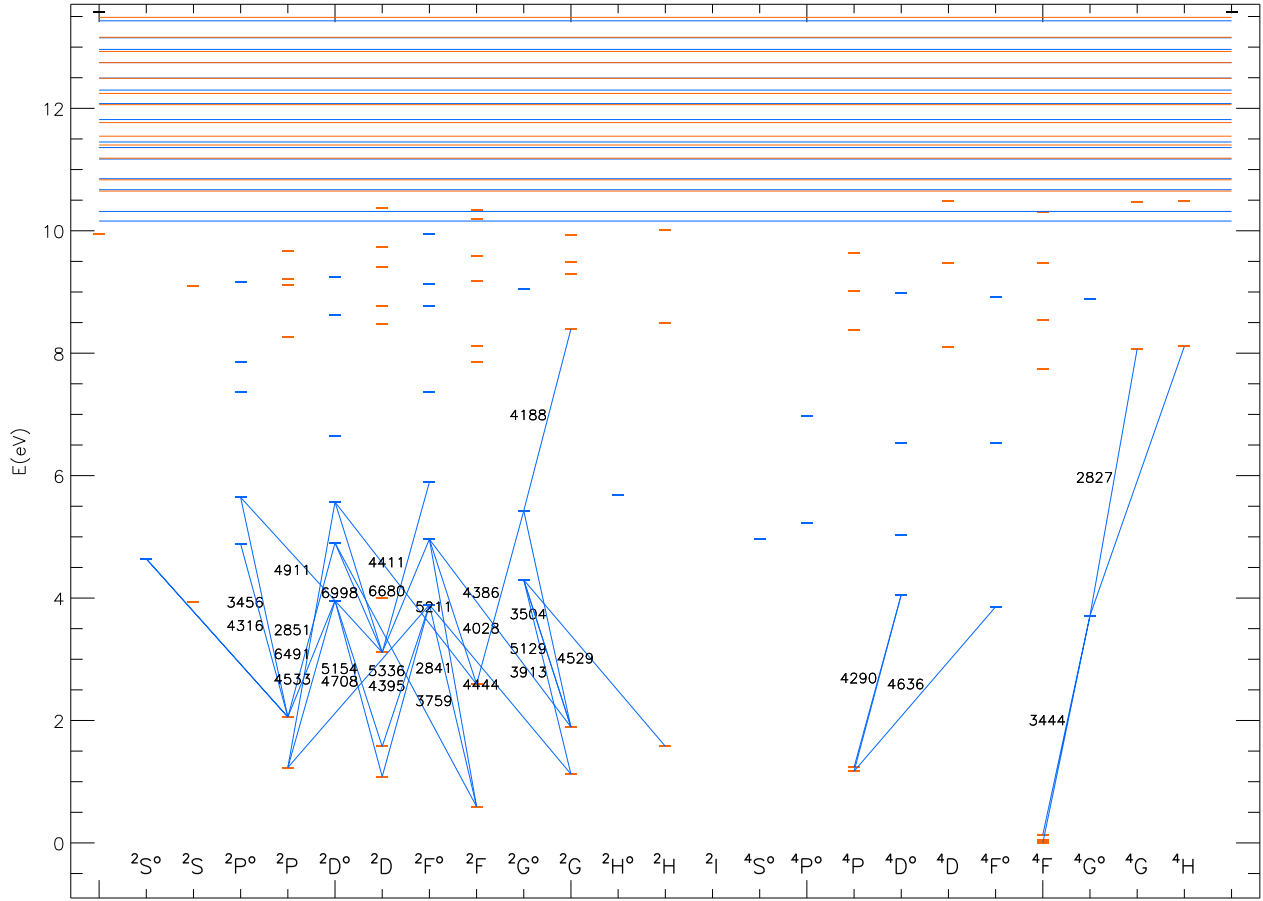}
	\caption{The same as in Fig.~\ref{ti1} for Ti\ii. The ionisation threshold for Ti\ii~is~13.57~eV. }
	\label{ti2} 
\end{figure*}

\underline{Energy levels.} Titanium is almost completely ionised throughout the atmosphere of stars with effective temperatures above 4500~K. For example, the ratio ${ \rm N_{Ti~II}/N_{Ti~I} \simeq 10^2}$ throughout the solar atmosphere. 
Such minority species as Ti\ione\ are particularly sensitive to NLTE effects because any small deviation in the intensity of ionising radiation from the Plank function strongly changes their population. 
For accurate calculations of the SE  we include in our model atom high-excitation levels of Ti\ione\ and Ti\ii, which establish collisional coupling of Ti\ione\ and Ti\ii\ levels near the continuum to the ground states of Ti\ii\ and Ti\iii, respectively. 
\citet{mash_fe} included high-excitation levels of Fe\ione\ in their Fe\ione--\ii\ model atom, and found that the SE of iron changed substantially by achieving close collisional coupling of the Fe\ione\ levels near the continuum to the ground state of Fe\ii. 
Our model atom of titanium (Fig.~\ref{ti1}, \ref{ti2}) is constructed using not only all the known energy levels from NIST \citep{NIST08}, but also the predicted levels from atomic structure calculation of R.~Kurucz ($http://kurucz.harvard.edu/atoms.html$). 
The measured levels of Ti\ione\ with the excitation energy \eexc~$\le$~6~eV belong to 175 terms.
Neglecting their fine structure, except for the ground state of Ti\ione, we obtain 177 levels in the model atom.
The predicted  and measured levels below the threshold, in total 3500 ones with \eexc~$\ge$~6~eV, with common parity and close energies were combined whenever the energy separation is smaller than $\Delta E$~=~0.1~eV.
This makes up 17 super-levels. 

For Ti\ii\ we use the experimental energy levels belonging to 89 terms with \eexc up to 10.5~eV. The fine structure is neglected, except for the ground state of Ti\ii. 
The  1800 high excitation levels with  $10.5 \le$~\eexc~$\le 13.6$~eV are used to make up 28 super-levels.
The ground state of Ti~\iii\ completes the system of levels in the model atom.

\underline{Radiative bound-bound (b-b) transitions.} In total, 7929 and 3104 allowed transitions of Ti\ione\ and Ti\ii, respectively, occur in our final model atom. Their average f-values are calculated using the data from R.~Kurucz database. 
We compared predicted gf-values with accurate laboratory data for about 900 transitions of Ti\ione\ \citep{Lawler2013_ti1} and found a systematic shift to be minor, with an average difference of log~gf$_{lab}$~--~log~gf$_{Kurucz}$~=~$-0.05 \pm$~0.28. An advantage of the Kurucz's predicted gf-values is their completeness that is of extreme importance for the statistical equilibrium  calculations.
For the transitions involving the superlevels the total gf-value was calculated as a sum of gf of individual transitions $g f_{tot} = \sum_{i, j}^{} (g_i f_{i, j})$, i~=~1,...,~N$_l$, j~=~1,..., N$_u$, where N$_l$ and N$_u$ are numbers of individual levels, which form a lower and upper superlevel, respectively.
Radiative rates were computed  using the Voigt profiles for transitions with  $f_{ij} \ge 0.10$ and  1800~\AA~$ \le \lambda \le$~4000~\AA\ and the Doppler profiles for the remaining ones. The transitions with $f_{ij} \le 10^{-8}$ were treated as forbidden ones. 

\begin{figure*}
	\centering
	\includegraphics[width=80mm]{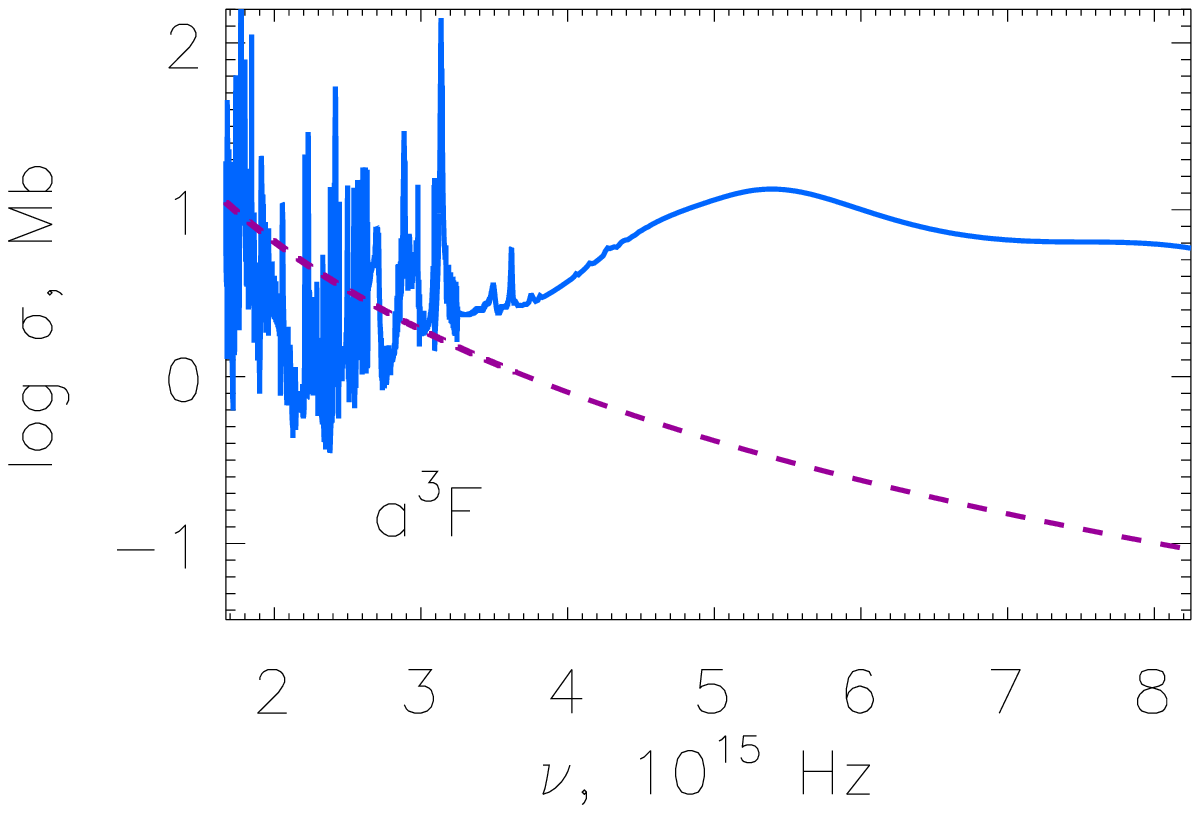}
	\includegraphics[width=80mm]{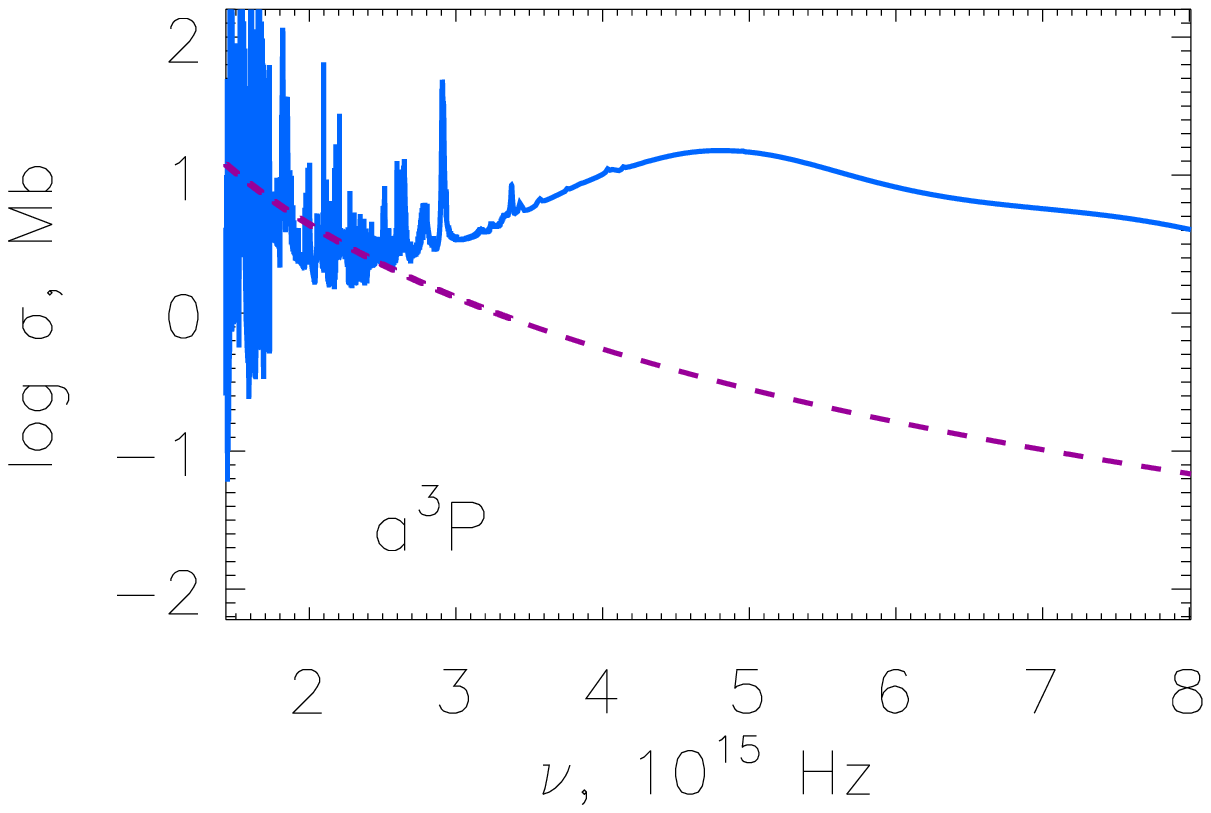}	
	\caption{Photoionisation cross sections for the ground state of Ti\ione\ (left panel) and the low-excitation level a3P  (right panel) from quantum mechanical calculations \citep[][continuous curve]{Nahar2015}, and computed in the hydrogenic approximation (dashed curve).} 
	\label{pic_ti1} 
\end{figure*}

\begin{figure*}
	\centering
	\includegraphics[width=80mm]{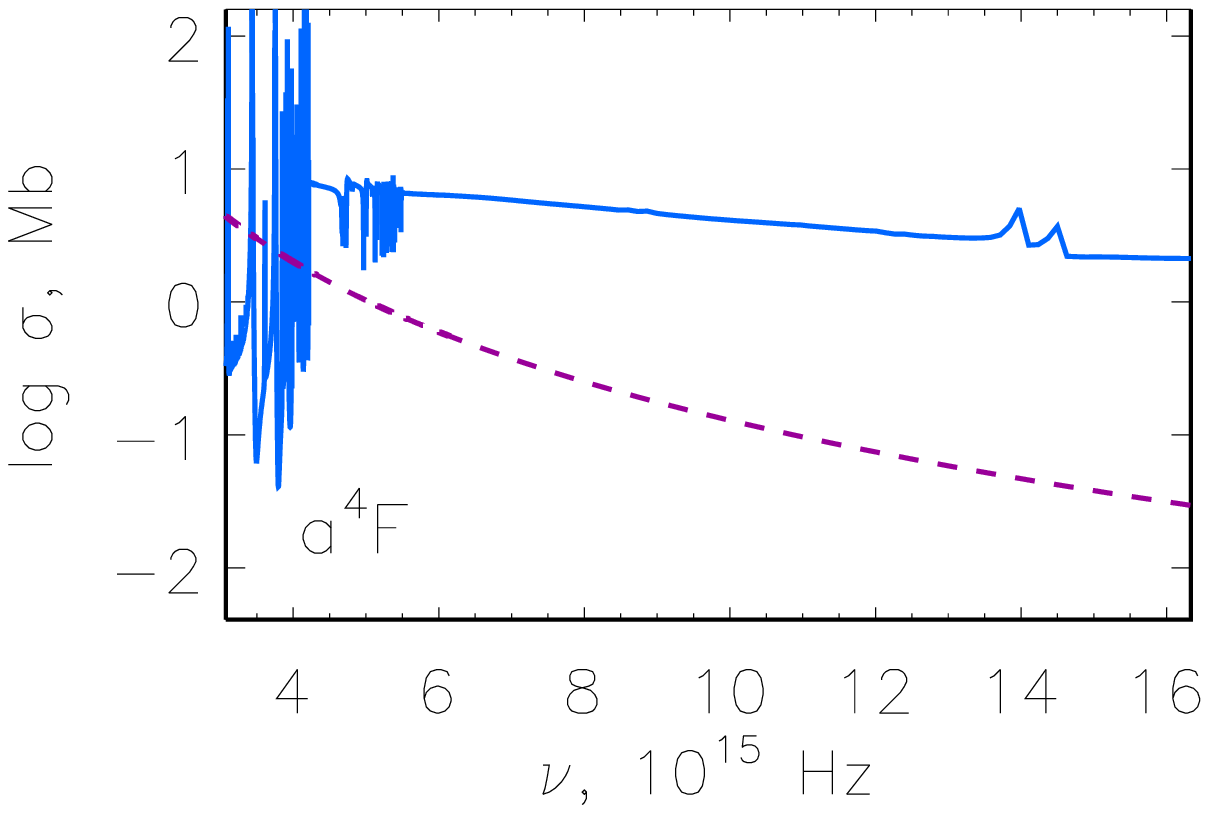}
	\includegraphics[width=80mm]{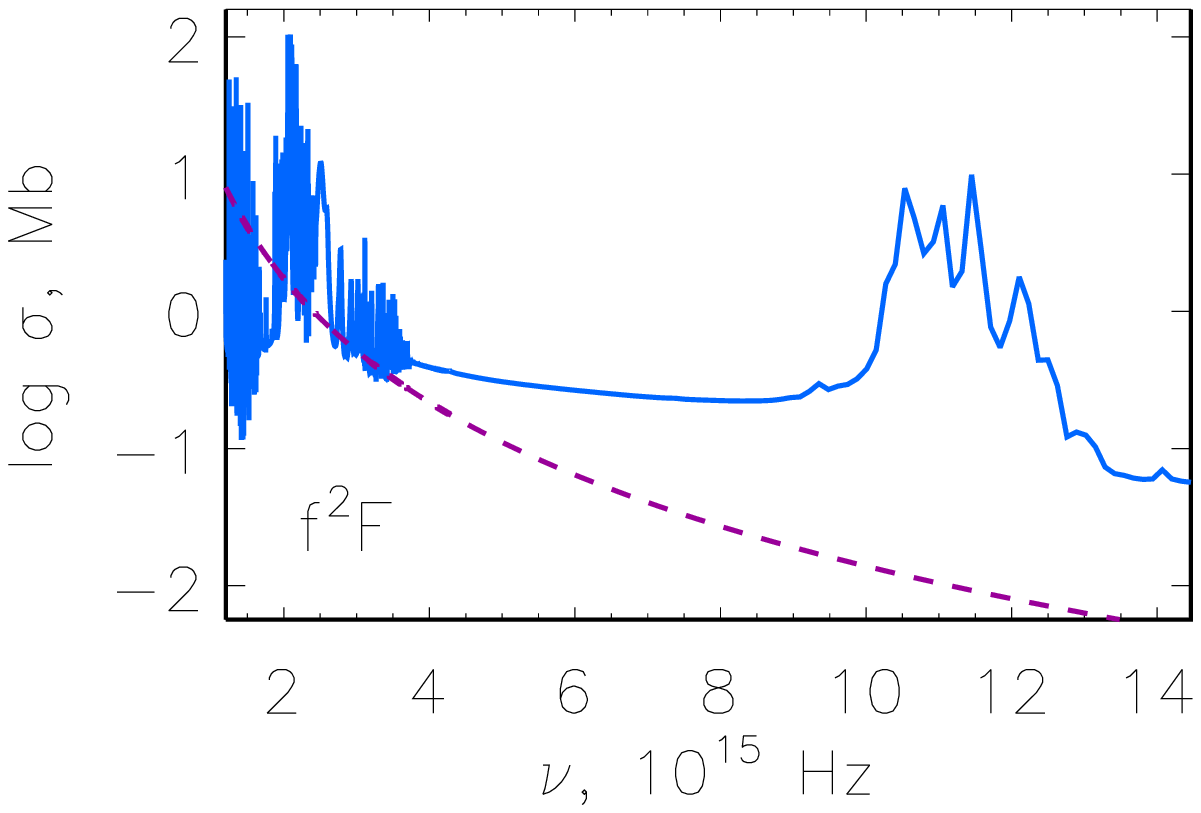}			
	\caption{Photoionisation cross sections for the ground state of Ti\ii\ (left panel) and the level f2F (right panel) from quantum mechanical calculations (K.~Butler, continuous curve) and computed in the hydrogenic approximation (dashed curve).} 
	\label{pic_ti2} 
\end{figure*}

\underline{Radiative bound-free (b-f) transitions.} 
For 115 terms of Ti\ione\ with \eexc~$\le$~5.5~eV we use photoionisation cross-sections from calculations of \citet{Nahar2015}, based on the close-coupling R-matrix method, and for 78 terms of Ti\ii\ with \eexc~$\le$~10.0~eV we use the data from quantum-mechanical calculations of Keith~Butler (private communication). For the remaining high-excitation levels we assume a hydrogenic approximation with using an effective principle quantum number.  
We compare the quantum-mechanical photoionisation cross-sections with the hydrogenic ones for selected levels of Ti\ione\ and Ti\ii\ in Fig.~\ref{pic_ti1} and \ref{pic_ti2}, respectively. 
For each level the hydrogenic cross-sections fit, on average, the quantum-mechanical ones near the ionisation threshold. The difference at frequencies higher than 3.29~$10^{15}$~Hz ($\lambda \le$~912 \AA) weakly affects the photoionisation rate because of small flux in this spectral range in the investigated stellar atmospheres. 

\underline{Collisional transitions.}
All levels in our model atom are coupled via collisional excitation and ionisation by electrons and by neutral hydrogen atoms.
Our calculations of collisional rates rely on the theoretical approximations because no accurate experimental or theoretical data are available.
For electron-impact excitation we use the formula of \citet{Reg1962} for the allowed transitions and the formula from \citet{WA1948_cbb} with a collision strength of 1.0 for the radiatively forbidden transitions.
Ionisation by electronic collisions is calculated from the \cite{Seaton1962} approximation using the threshold photoionisation cross-section. 

For collisions with H\ione\  atoms, we employ the formula of \citet{SteenbockHolweger1984} based on theory of \citet{Drawin1968,Drawin1969} for allowed b-b and b-f transitions and, following \citet{Takeda1994}, a simple relation between hydrogen and electron collisional rates, C$_H = C_e \sqrt{(m_e/m_H)}N_H/N_e$, for forbidden transitions. 
Due to the Drawin formula provides order-of-magnitude estimates, we perform the NLTE calculations using a scaling factor \kH=0.1, 0.5 and 1, and constrain its magnitude empirically from analysis of metal-poor stars.

The nearly resonance charge exchange reaction (CER)  H$^+$ + Ti\ii\ $\leftrightarrow$ H\ione\ + Ti\iii\ takes place because the ionisation thresholds for Ti\ii\ and H\ione\ are 13.57~eV and 13.60~eV, respectively. There are no literature data on cross-sections for this process. 
In order to inspect an influence of CER on the statistical equilibrium of titanium, we assumed that the analytic fit deduced by \citet{Arnaud1985} for O\ione\ can also be applied to Ti\ii, because the ionisation threshold for O\ione\ is close to that for Ti\ii\ and amounts to 13.62~eV. 
Test calculations for A-type stars showed that the CER makes the populations of the ground states of Ti~\iii\ and  Ti\ii\ to be in thermodynamic equilibrium, nevertheless, no change in the NLTE abundances from lines of Ti\ii\ was found.  For stars with \teff~$\le$~9000~K the CER weakly affects the SE because of small fraction of Ti\iii.

\subsection{Programs and model atmospheres}

The coupled radiative transfer and SE equations were solved with a revised version of the  \textsc {detail} code by \citet{detail}. 
The opacity package of the \textsc {detail} code was updated as described by  \citet{2011JPhCS.328a2015P} and \citet{mash_fe}, by including the quasi-molecular Ly$_{\alpha}$ satellites following the implementation by \citet{CastelliKurucz2001} of the \citet{Allard1998} theory and using the Opacity Project  \citep[see][for a general review]{Seaton1994} photoionisation cross-sections for the calculations of b-f absorption of C\ione, N\ione, O\ione, Mg\ione, Si\ione, Al\ione, Ca\ione, and Fe\ione. In addition to the continuous background opacity, the line opacity introduced by H\ione\ and metal lines was taken into account by explicitly including it in solving the radiation transfer. The metal line list was extracted from the \citet{Kurucz1994} compilation and the VALD database \citep{vald}. 	
The pre-calculated departure coefficients were then used by \textsc {synthV\_NLTE} code updated in \citet{ryab2015}, and based on \citet{Tsymbal1996} to compute the theoretical synthetic spectra. 
The integration of the \textsc {synthV\_NLTE} code in the \textsc {idl binmag3} code by O. Kochukhov\footnote{http://www.astro.uu.se/$\sim$oleg/download.html} allows us to obtain the best fit to the observed line profiles with the NLTE effects taken into account.  

Throughout this study, the element abundance is determined from line profile fitting.
For late type stars we used classical plane-parallel model atmospheres from the \textsc {marcs} model grid \citep{MARCS}, which were interpolated for given \teff, \lgg, and [Fe/H] using a FORTRAN-based routine written by Thomas Masseron\footnote{{\tt http://marcs.astro.uu.se/software.php}}.
For A-B type stars the model atmospheres were calculated under the LTE assumption with the code \textsc {LLmodels} \citep{LLmodels}.

For each star the line list includes unblended lines of various strength ($EW\le$~150~m\AA, where $EW$ is the line equivalent width) and excitation energies. The full list of the lines is presented in Table~\ref{atomic} along with the transition information, gf-value, excitation energy, and damping constants (log~$\gamma_{rad}$, log~$\gamma_4/N_e$, log~$\gamma_6/N_H$ at 10000 K).
The line list was extracted from the VALD database \citep{vald,vald2015}. The adopted oscillator strengths for most lines of both ions were measured by a common method \citep[][-- Wisconsin data]{Lawler2013_ti1,Wood2013_ti2}, and, hence, represent homogeneous set of gf-values. 

\begin{table*}
	\caption{The list of Ti\ione\ and Ti\ii\ lines with the adopted atomic data. This table is available in its entirety in a machine-readable form in Sect.\ref{apend}. A portion is shown here for guidance regarding its form and content.}
	\renewcommand\arraystretch{1.1}
	\vspace{2mm}
	\begin{large}
		\centering
		\label{atomic}

		\begin{tabular}{|l|c|c|c|c|c|c|}
			\hline
			\textbf{$\lambda$, \AA} & \multicolumn{1}{c}{\textbf{ \eexc, eV}} & \multicolumn{1}{c}{\textbf{  log gf}} & transition & \multicolumn{1}{c}{\textbf{log $\gamma_{rad}$}} & \multicolumn{1}{c}{\textbf{log $\gamma_4/N_e$}} & \multicolumn{1}{c}{\textbf{log  $\gamma_6/N_H $}} \\
			\hline
			\multicolumn{7}{l}{Ti\ione} \\
			4008.927 &  0.021 &     -1.000 &   3a3F --   y3F &      8.000 &     -6.080 &     -7.750  \\                    
			4060.262 &  1.052 &     -0.690 &   a3P  --   x3P &      8.050 &     -6.050 &     -7.646 \\                    
			4287.403 &  0.836 &     -0.370 &   a5F  --   x5D &      8.230 &     -6.010 &     -7.570   \\                    
			4449.143 &  1.886 &      0.470 &   a3G  --   v3G &      8.120 &     -5.560 &     -7.579   \\                        
			4453.699 &  1.872 &      0.100 &   a3G  --   v3G &      8.110 &     -4.970 &     -7.582   \\                    
			4512.733 &  0.836 &     -0.400 &   a5F  --   y5F &      8.130 &     -5.120 &     -7.593   \\                    
			4533.240 &  0.848 &      0.540 &   a5F  --   y5F &      8.130 &     -5.120 &     -7.593  \\                  
			4534.776 &  0.836 &      0.350 &   a5F  --   y5F &      8.130 &     -5.280 &     -7.596  \\
			4548.763 &  0.826 &     -0.280 &   a5F  --   y5F &      8.130 &     -5.410 &     -7.598  \\
			4555.484 &  0.848 &     -0.400 &   a5F  --   y5F &      8.130 &     -5.280 &     -7.596  \\    
			\multicolumn{7}{l}{...} \\
			\hline
		\end{tabular}
	\end{large}
\end{table*}

\subsection{Statistical equilibrium of Ti\ione\--\ii}

\begin{figure*}
	\centering
	\includegraphics[width=80mm]{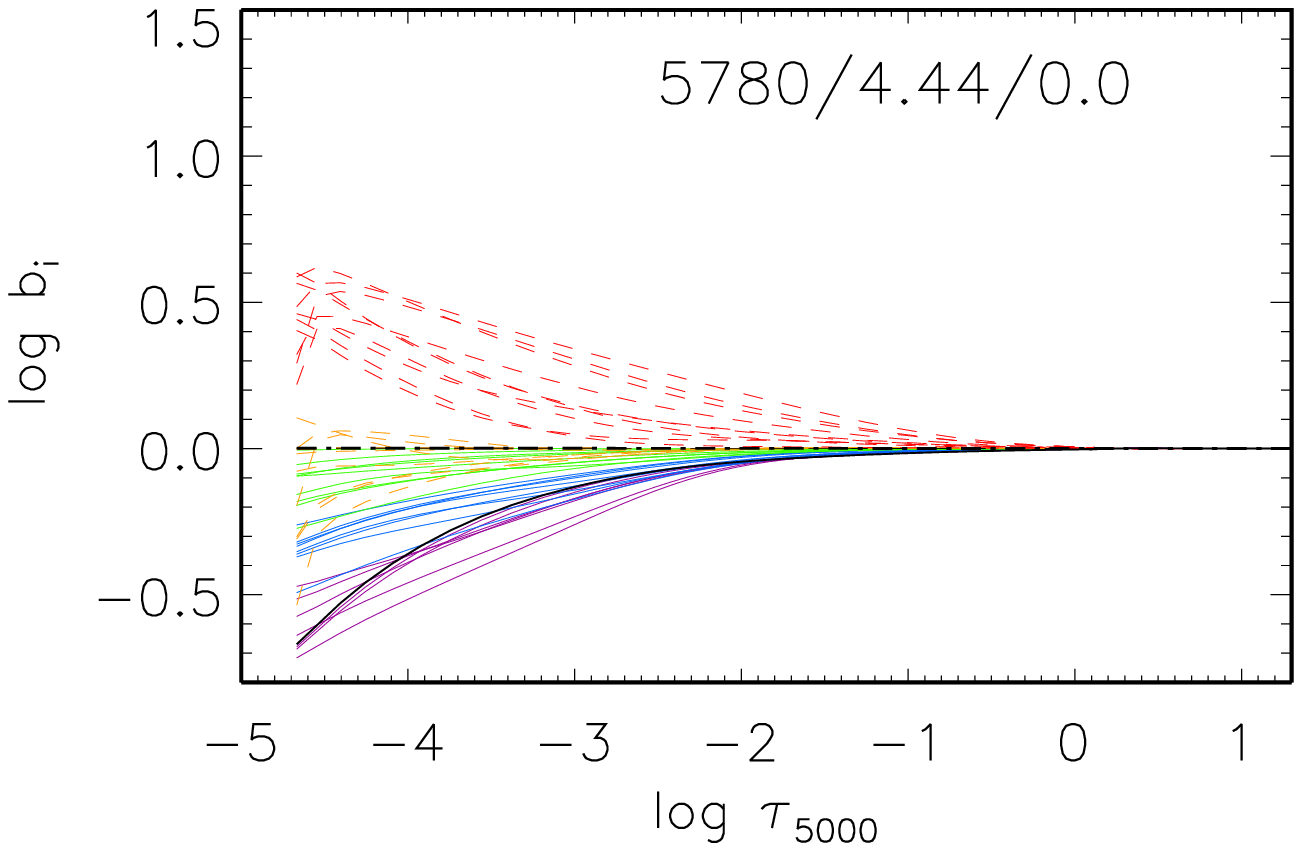}
	\includegraphics[width=80mm]{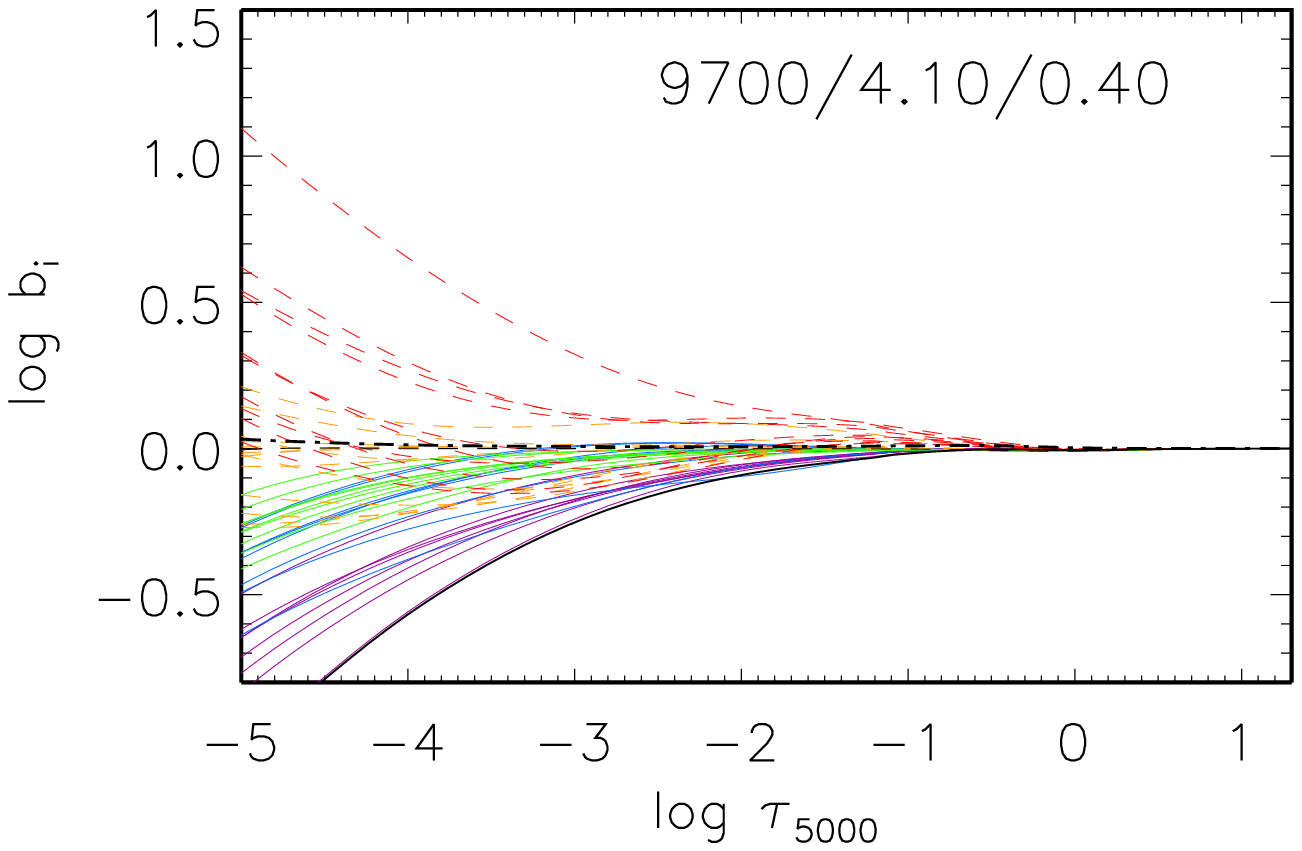}
	\includegraphics[width=80mm]{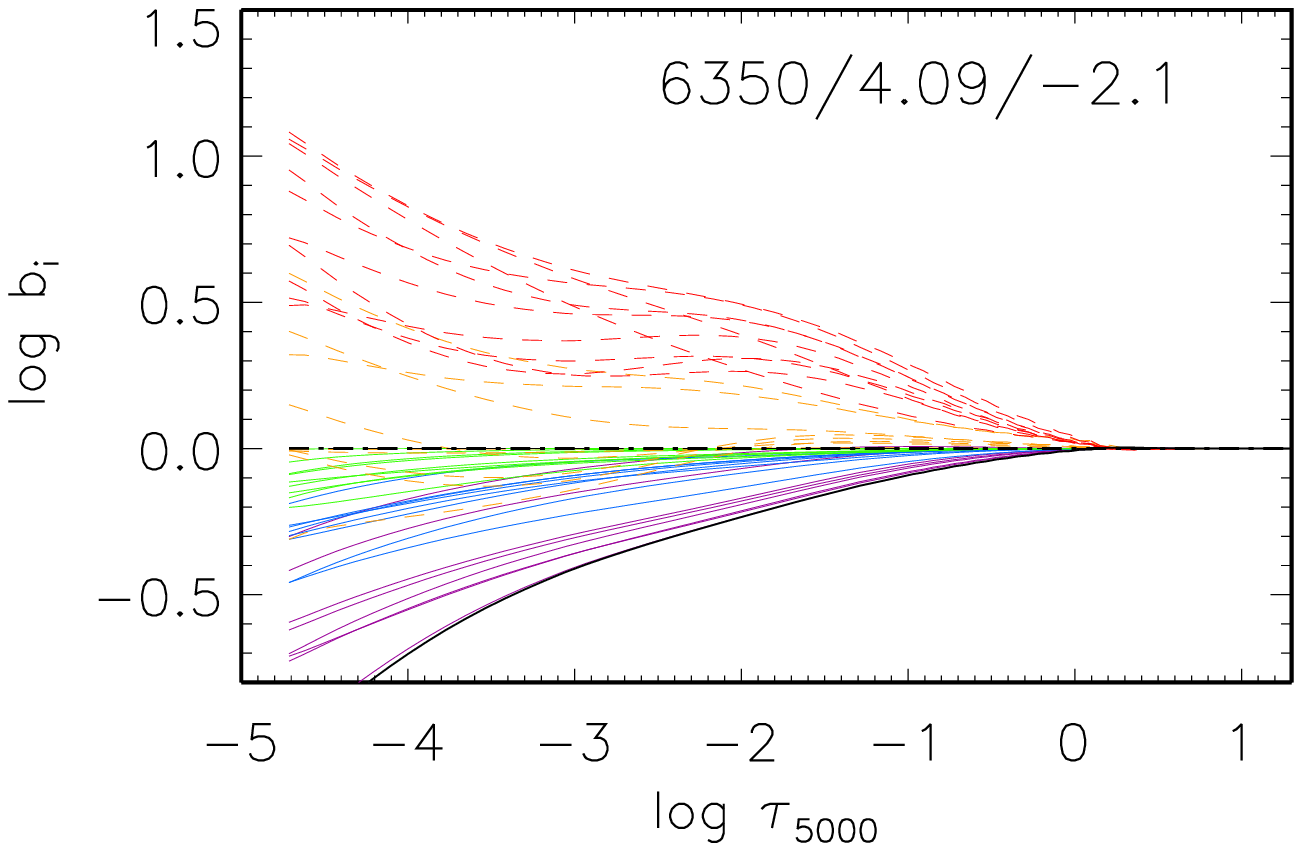}
	\includegraphics[width=80mm]{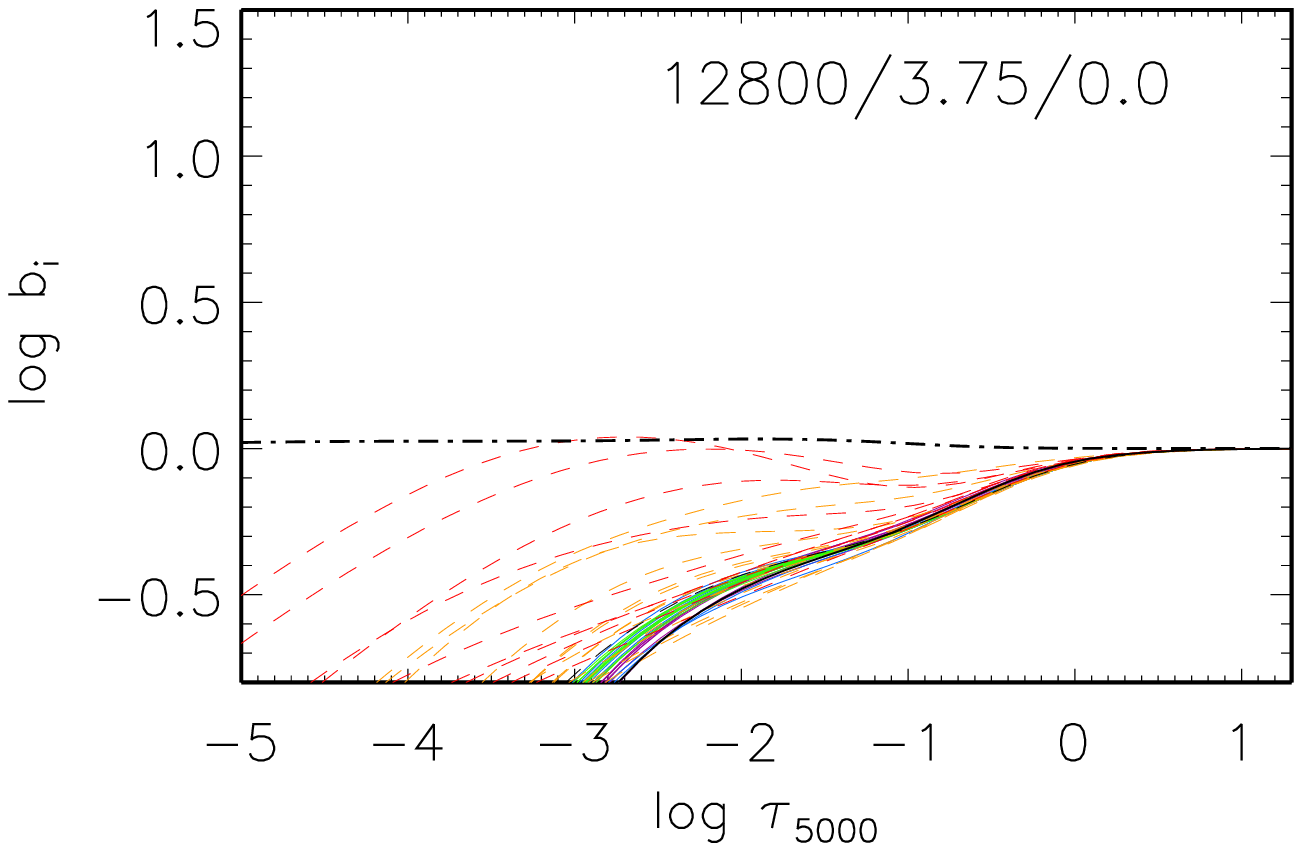}		
	\caption{Departure coefficients for the levels of Ti\ione\ (continuous curves) and Ti\ii\ (dotted curves) and the ground state of Ti~III (dash-dotted curve) in different model atmospheres. For each model atmospheric parameters \teff / \lgg/ [Fe/H] are indicated.}
	\label{depart} 
\end{figure*}

In this section, we consider the NLTE effects for Ti\ione--\ii\ in various model atmospheres.
The deviations from LTE in level populations are characterized by the departure coefficients b$_{i}$~=~n$^{NLTE}_i$/n$^{LTE}_i$, where n$^{NLTE}_i$ and n$^{LTE}_i$ are the statistical equilibrium and thermal (Saha-Boltzmann) number densities, respectively.
%
%
The departure coefficients for the selected levels of Ti\ione, Ti\ii\ and the ground state of Ti~\iii\ in the model atmospheres 5777/4.44/0, 6350/4.09/$-2.15$, 9700/4.1/0.4 and 12800/3.75/0  are presented in Fig.~\ref{depart}.
All the levels retain their LTE populations in deep atmospheric layers below log$\tau_{5000}$~=~0. 
In the higher atmospheric layers a total number density of Ti\ione\ is lower compared with the TE value. The overionisation is caused by superthermal radiation of non-local origin below the thresholds of the low excitation levels of Ti\ione. In the atmospheres, where Ti\ii\ is the majority species, collisional recombinations to the Ti\ione\ high-excitation levels followed by cascades of spontaneous transitions tend to compensate a depopulation of the lower levels of Ti\ione. However, this process can not prevent the overionisation. 
High superlevels of Ti\ione\ are collisionally coupled to the ground state of Ti\ii. NLTE leads to weakened lines of Ti\ione\ compared to their LTE strengths.

High levels of Ti\ii\ are overpopulated via radiative pumping transitions from the low excitation levels. The NLTE effects for Ti\ii\ are small in cool atmospheres. 
In the models 5780/4.44/0.0 and 6350/4.09/$-2.1$ a behaviour of the departure coefficients is qualitatively similar. However a magnitude of the NLTE effects grows towards higher \teff\ and lower \lgg\ and [Fe/H]. 
In the models representing atmospheres of A-type stars high levels of Ti\ii\ retain their LTE populations inward log$\tau_{5000}$~=~$-1.5$, and become underpopulated in the higher atmospheric layers. This results in strengthening the Ti\ii\ line cores formed in the uppermost layers compared with LTE. 
In the hottest model atmosphere 12800/3.75/0.0 Ti~\iii\ becomes the majority species, while the levels of Ti\ii\ are underpopulated beginning at log~$\tau \simeq$~0.5. Overionisation of Ti\ii\ results in weakened Ti\ii\ lines.   

\section{Observations and stellar atmosphere parameters}
\label{obspar}

Our sample includes the Sun and 24 well-studied stars. They are listed in Table~\ref{obs}.
Atmospheric parameters (\teff, \lgg, [Fe/H], \vt) were either determined in our earlier studies or taken from the literature.  These parameters were derived by several independent methods, which gave consistent results. 
Our hot stellar sample consists of A and late B stars, which do not reveal pulsation activity, chemical stratification and magnetic field.
For Sirius, $\pi$~Cet, 21 Peg,  HD~32115, HD~37594,  HD~73666, HD~145788 atmospheric parameters were derived by common method, based on multicolour photometry, analysis of hydrogen Balmer lines and metal lines in high resolution spectra and comparison of spectrophotometric data with theoretical flux (see Table~\ref{obs} for the references). 
For HD~72660 the parameters 9700/4.10/0.45/1.8 were derived by fitting the 4400-5200~\AA\ and 6400-6700~\AA\ spectral regions with \textsc{sme} (Spectroscopy Made Easy) program package \citep{1996AAS..118..595V}. We used medium-resolution spectrum of HD~72660 extracted from ELODIE archive\footnote[1]{http://atlas.obs-hp.fr/elodie/}.   
\textsc{sme} was tested for Sirius, $\pi$~Cet, 21~Peg, and HD~32115 by \citet{2015ASPC..494..308R} where the authors derived practically the same parameters as adopted in the present paper. Atmospheric parameters of HD~72660 agree with the results of \citet{Lemke1989}, who derived \teff /\lgg~=~9770/4.0 from photometry and H$_{\beta}$, and \citet{2009A&A...503..973L}, who derived 9650/4.05.

Each cool star of the sample has photometric \teff\  and  \lgg\ based on the Hipparcos parallax. 
We checked in advance whether an ionisation equilibrium between Fe\ione\ and Fe\ii\ is fulfilled in NLTE when using non-spectroscopic parameters. The iron abundances obtained from the lines of Fe\ione\ and Fe\ii\ in dwarfs agree within 0.05~dex in NLTE, when using \kH~=~0.5 \citep{lick}. To confirm the adopted parameters, we checked them with evolutionary tracks and derived reasonable masses and ages. 
Our sample also includes the most metal-poor giant, HD~122563 ([Fe/H]~=~$-2.56$), with the accurate Hipparcos parallax available. The effective temperature of HD~122563 was determined by \citet{creevey2012} based on angular diameter measurements.

\begin{table*}
	\caption{Stellar atmosphere parameters and characteristics of the observed spectra.}
	\renewcommand\arraystretch{1.1}
	\vspace{2mm}
		\centering
		\vspace{5mm}
		\label{obs}

		\begin{tabular}{|l|c|c|c|c|c|c|c|c|}
			\hline
			Star & \teff , & $\log g$ & [Fe/H] & \vt, &   Ref. &$\lambda/\Delta\lambda$, & $S/N >$ & source\\
			&          K &          &        & \kms  &           &    10$^3$                   &         &         \\
			\hline
			Sun   &  5777 & 4.44 &   0.0 & 0.9  & --  & 300& 300& KPNO84 \\
			HD~24289 &  5980 & 3.71 & --1.94 & 1.1 &  S15 & 60 & 110 & S15 \\
			HD~64090 & 5400 & 4.70 &   --1.73 & 0.7   & S15 & 60 & 280 & S15 \\ 
			HD~74000  &  6225 & 4.13 &  --1.97 & 1.3 &    S15 & 60 & 140 & S15 \\
			HD~84937 &  6350 & 4.09 &   --2.16 &  1.7 &   S15 &  80 & 200 & UVESPOP$^1$  \\
			HD~94028 & 5970 & 4.33 &   --1.47 & 1.3 &    S15  & 60 & 120 & S15 \\ 
			HD~103095 & 5130 & 4.66 &   --1.26 &  0.9 &   S15 &  60 & 200 & FOCES$^2$ \\
			HD~108177 & 6100 & 4.22 &   --1.67 &  1.1 &   S15 &  60 & 60 & S15 \\
			HD~140283 &  5780 & 3.70 &   --2.46 & 1.6 &  S15 & 80 & 200 & UVESPOP  \\
			BD--4$^\circ$ 3208     & 6390 & 4.08 &   --2.20 & 1.4   & S15  & 80 & 200 & UVESPOP   \\
			BD--13$^\circ$ 3442   &  6400 & 3.95 &  --2.62 & 1.4    & S15   & 60 & 100 & S15 \\
			BD+7$^\circ$ 4841  &  6130 & 4.15 &   --1.46 & 1.3    &  S15  & 120 & 150 & S15 \\
			BD+9$^\circ$ 0352  &  6150 & 4.25 &   --2.09 & 1.3    &  S15  & 120 & 160 & S15 \\
			BD+24$^\circ$ 1676  & 6210 & 3.90 &  --2.44 & 1.5 &   S15 & 60 & 90 & S15 \\
			BD+29$^\circ$ 2091  &  5860 & 4.67 &   --1.91 & 0.8   & S15  & 60 & 80 & S15 \\
			BD+66$^\circ$ 0268  &  5300 & 4.72 &   --2.06 & 0.6   &  S15  & 60 & 110 & S15  \\
			G~090--003          &  6010 & 3.90 &   --2.04 & 1.3   &  S15  & 60 & 100 & S15 \\
			HD~122563 &  4600 & 1.60 &   --2.60 & 2.0 & M11 & 80 & 200  & UVESPOP \\
			HD~32115 &  7250 & 4.20 &   0.0 & 2.3 & F11  & 60 & 490 & F11 \\
			HD~37594  &  7150 & 4.20 &  -0.30 & 2.5 & F11   & 60 & 535 & F11 \\                 
			HD~72660  & 9700 & 4.10 & 0.45 & 1.8 & this study  & 30 & 150 & STIS$^3$,  L98 \\
			HD~73666 & 9380 & 3.78 & 0.10 & 1.8 & F07, F10  & 65 & 660 &  F07 \\
			HD~145788 & 9750 & 3.70 & 0.0 & 1.3 & F09 & 115 & 200 &  F09 \\
			HD~209459 & 10400 & 3.55 & 0.0 & 0.5 &  F09 & 120 & 700 &  F09 \\
			(21 Peg) &     &    &  &&&&&                  \\
			HD~48915 & 9850 & 4.30 &   0.4 & 1.8$^4$ & H93  & 70 & 500 & F95 \\
			(Sirius) &     &    &   &&&&&                  \\
			HD~17081 & 12800 & 3.75 & 0.0 & 1.0 &   F09 & 65 & 200 &  F09 \\
			($\pi$ Cet) &     &    &         &&&&&            \\
			\hline
			\multicolumn{9}{p{.8\textwidth}}{$^1$~\cite{Bagnulo2003}, $^2$~K.~Fuhrmann, private communication, $^3$~J.~Landstreet, private communication, $^4$~\cite{Sitnova2013}, KPNO84 = \cite{kurucz84}, S15 = \cite{lick},   M11 = \cite{mash_fe}, H93 = \cite{hill1993}, F95 = \cite{Furenlid1995}, F07 = \cite{fossati07}, F09 = \cite{fossati09}, F10 = \cite{fossati10}, F11 = \cite{fossati11}, L98 = \cite{Landstreet1998}.}\\		
			\hline			
		\end{tabular}
\end{table*}

\section{Analysis of Ti\ione\ and Ti\ii\ lines in A-B-type stars}
\label{hots}

A-B type stars are suitable for testing the treated model atom because the deviations from LTE are large for both Ti\ione\ and Ti\ii\ and poorly known inelastic collisions with hydrogen atoms do not or weakly affect the SE.
For example, in the model 7170/4.20/$-0.30$ the use of \kH~=~0 and 0.5 leads to a maximal abundance difference of 0.02~dex and 0.01~dex for individual lines of Ti\ione\ and Ti\ii, respectively.

The lines of two ionisation stages are observed in  HD~32115, HD~37594, HD~73666, and HD~72660. In spectra of Sirius, 21~Peg, $\pi$~Ceti, and HD~145788 only the lines of Ti\ii\ can be detected. 
For each star at least 6 lines were used to derive the titanium abundance.
The LTE and NLTE abundances are given in Table~\ref{abund}. In NLTE, the abundance from Ti\ione\ lines increases by 0.05~dex to 0.14~dex for different stars. In contrast, NLTE leads to up to 0.12~dex lower abundance from the lines of Ti\ii.     
An exception is the late B star $\pi$ Cet, where NLTE leads to line weakening and to higher titanium abundance compared with LTE. From the eleven lines of  Ti\ii\ we derived  log~A(Ti)~=~$-7.41 \pm$0.09~dex and log~A(Ti)~=~$-7.14 \pm$0.08~dex in LTE and NLTE, respectively.
Hereafter, the statistical abundance error is the dispersion in the single line measurements:
$\sigma = \sqrt{\Sigma (x - x_i )^2 /(N - 1)}$, where N is the total number of lines used, x is their mean abundance, $x_i$ is the  abundance of each individual line.
In LTE for four our stars the abundance difference Ti\ione--Ti\ii ranges between $-0.22$~dex and $-0.09$~dex, while in NLTE Ti\ione--Ti\ii decreases in absolute value and does not exceed 0.07~dex for each of the four stars.

\begin{figure}
	\centering
	\includegraphics[width=80mm]{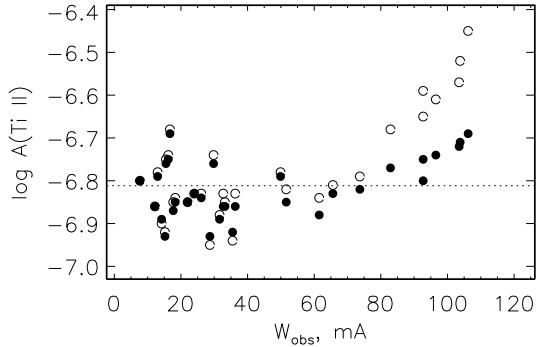}	
	\caption{NLTE (filled circles) and LTE (open circles) abundances from the lines of Ti\ii\ in HD~145788 as a function of equivalent width.} 
	\label{hd145788}
\end{figure}

For the A-type stars the LTE abundances from strong lines of Ti\ii\ are higher than those from the weak lines (see Fig.~\ref{hd145788} for HD~145788). Such a behavior can be wrongly interpreted as an underestimation of a microturbulent velocity. 
For example, to derive consistent LTE abundances from different lines of Ti\ii\ in HD~145788 one needs to adopt a microturbulent velocity of \vt~=~1.8~\kms, while \vt~=~1.3~\kms\  was found by \citet{fossati09} from lines of Fe\ii.
We show that a discrepancy between strong and weak lines vanishes in NLTE.
This is because the strong lines are more affected by NLTE  compared with the weak lines. 
For example, in HD~145788, the cores of the Ti\ii\ lines with $EW\sim$~100~m\AA\ form at the optical depth log$\tau_{5000} \simeq$~--2.5, and their NLTE abundance corrections reach --0.24~dex. 
For the Ti\ii\ lines with $EW\le$~70~m\AA\ the NLTE abundance corrections do not exceed few hundredths in absolute value. We do not recommend to apply the Ti\ii\ lines with $EW\ge$~70~m\AA\ for abundance determination under the LTE assumption. For A-B type stars NLTE leads to significant decrease of line-to-line scatter compared to LTE (Table~\ref{abund}).

We checked effects of the use of accurate photoionisation cross-sections by \citet{Nahar2015} and K. Butler instead of the hydrogenic approximation.
Using quantum-mechanical cross-sections for Ti\ione\ leads to increasing the photoionisation rates and the deviations from LTE. For example, the NLTE abundance corrections for Ti\ione\ lines increase by 0.01--0.02~dex in the model 9700/4.10/0.4/1.8.
In the atmospheres with \teff~$\leq$ 10500~K the NLTE abundances derived from the Ti\ii\ lines do not change significantly,  when using either accurate or hydrogenic cross-sections. This is due to the fact that mechanism of deviations from LTE for Ti\ii\ is not ruled by the bound-free transitions. 
For the hottest star of our sample, HD~17081 (B7~IV), where Ti\ii\ is affected by overionisation, we found that
using the accurate cross-sections leads to weakened NLTE effects for Ti\ii\ and 0.06~dex smaller NLTE abundance  compared with that calculated with the hydrogenic cross-sections.
Since we adopt the theoretical approximations to calculate electron collision rates, we perform the test calculations.
Test calculations with the model atmosphere 7250/4.20/0.0 show that a hundredfold decrease in electron collision rates results in a 0.05~dex increase in the NLTE abundance from Ti\ione,  and up to 0.06 dex decrease in NLTE abundance from the strongest lines of Ti\ii\ with EW of 150 m\AA.

Thus, analysis of the titanium lines in the hot stars gives an evidence for that our NLTE method gives reliable results. 

\begin{figure}
	\centering
	\includegraphics[width=80mm]{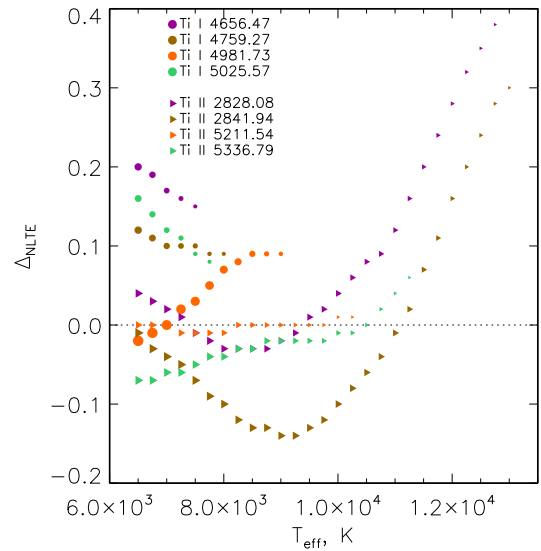}	
	\caption{NLTE abundance corrections for the selected lines of Ti\ione\ (circles) and Ti\ii\ (triangles) shown by different colours. A size of symbol represents an equivalent width of the corresponding line.} 
	\label{corrections_fig}
\end{figure}

For the 22 lines of Ti\ione\ and 82 lines of Ti\ii\ we calculated the NLTE abundance corrections in a grid of model atmospheres with \teff\ from 6500~K to 13000~K with a step of 250~K, \lgg~=~4, [Fe/H]~=~0 and \vt~=~2~\kms. For  lines of Ti\ione\ the NLTE abundance corrections are positive and vary between 0.0~dex to 0.20~dex (Fig.~\ref{corrections_fig}). For Ti\ii\ the NLTE abundance corrections are negative for  \teff\ $\leq$~10000~K and can be up to $-0.17$~dex. In the atmospheres with \teff~$\geq$~10000~K the lines of neutral titanium can not be detected, and the NLTE abundance corrections for lines of Ti\ii\ are positive and reach  0.37~dex. The data are available as on-line material (Table~\ref{corrections}).

\begin{table}
	\caption{NLTE abundance corrections and equivalent widths for the lines of Ti\ione\ and Ti\ii\ depending on \teff\ in the models with \lgg~=~4, [Fe/H]~=~0, and \vt~=~2~\kms. This table is available in its entirety in a machine-readable form in Sect.\ref{apend}. A portion is shown here for guidance regarding its form and content. If EW~=~$-1$ and \dnlte~=~$-1$, this means that EW~$<$~5~m\AA\ in a given model atmosphere.}
	\begin{large}
		\centering
		\label{corrections}
		\begin{tabular}{|r|r|r|r|r|}
			\hline
			\teff$_1$, K & \teff$_2$ & ... &  \teff$_{26}$ & \teff$_{27}$ \\
			\hline	
			EW$_1$, m\AA & EW$_2$ &  ... &  EW$_{26}$ & EW$_{27}$ \\
			\dnlte$_1$ & \dnlte$_2$ &  ... &  \dnlte$_{26}$ & \dnlte$_{27}$ \\				
			\hline	
			6500 & 6750 &  ... &  12750 & 13000 \\
			\multicolumn{5}{l}{ ... } \\
			\multicolumn{5}{l}{ 5210.3838 \AA\  Ti\ione\  \eexc =  0.048 eV  log gf = --0.820 } \\
			54  &  40  &   ... &      --1  &  --1 \\
			0.17 & 0.17 &  ... &   --1.00 & --1.00 \\
			\multicolumn{5}{l}{ ... } \\
			\multicolumn{5}{l}{ 4395.0308 \AA\  Ti\ii\  \eexc =  1.084 eV log gf =  --0.540 } \\
			176  & 169 &   ... &    15  &  12 \\
			--0.09 & --0.10 &  ... &   0.25 & 0.26	\\	
			\hline	
		\end{tabular}
	\end{large}
\end{table}

\section{Analysis of Ti\ione\ and Ti\ii\ lines in the reference late-type stars}
\label{cools}

\subsection{Ti\ione\ and Ti\ii\ lines in the solar spectrum}
We used 27 Ti\ione\ and 12 Ti\ii\ lines in the  solar flux spectrum \citep{kurucz84} to determine the LTE and NLTE abundances.
Under the LTE assumption we derived log~A$_{\rm TiI} = -7.11 \pm$~0.06 dex and log~A$_{\rm TiII} = -7.06 \pm$~0.04 dex from the lines of Ti\ione\ and Ti\ii, respectively. 
We calculated the NLTE abundances for \kH~=~0, 0.1, 0.5 and 1.0. 
Consistent within 0.03~dex abundances from  Ti\ione\ and Ti\ii\ were found in NLTE, independent of adopted \kH\ value (Table~\ref{lines5}).
This means that the solar analysis does not help to constrain \kH.
Solar titanium abundance averaged over Ti\ione\ and Ti\ii\ lines, log~A~=~$-7.09 \pm$0.06 (NLTE, \kH~=~1), agrees with the meteoritic value,  log~A~=~$-7.11 \pm$0.03~dex \citep{Lodders2009}.

The treated NLTE method was applied before publication to check the Ti\ione/Ti\ii\ ionisation equilibrium of 11 stars with 
5050~$\le$~\teff~$\le$~6600~K, 3.76~$\le$~\lgg~$\le$~4.47 and $-0.48 \le$~[Fe/H]~$\le 0.24$ \citep{ryab2015}.
For HD~49933 (6600/4.0/$-0.48$), the star with the largest deviations from LTE in the sample, the NLTE calculations provide consistent within the error bars the Ti\ione\ and Ti\ii\ based abundances independent of using either \kH~=~0.5 or 1.
For the studied stars the NLTE abundance difference Ti\ione--Ti\ii\ nowhere exceeds $-0.06$~dex.

\subsection{Ti\ione\ and Ti\ii\ lines in the metal-poor stars} 
\begin{figure*}
	\centering
	\includegraphics[width=80mm]{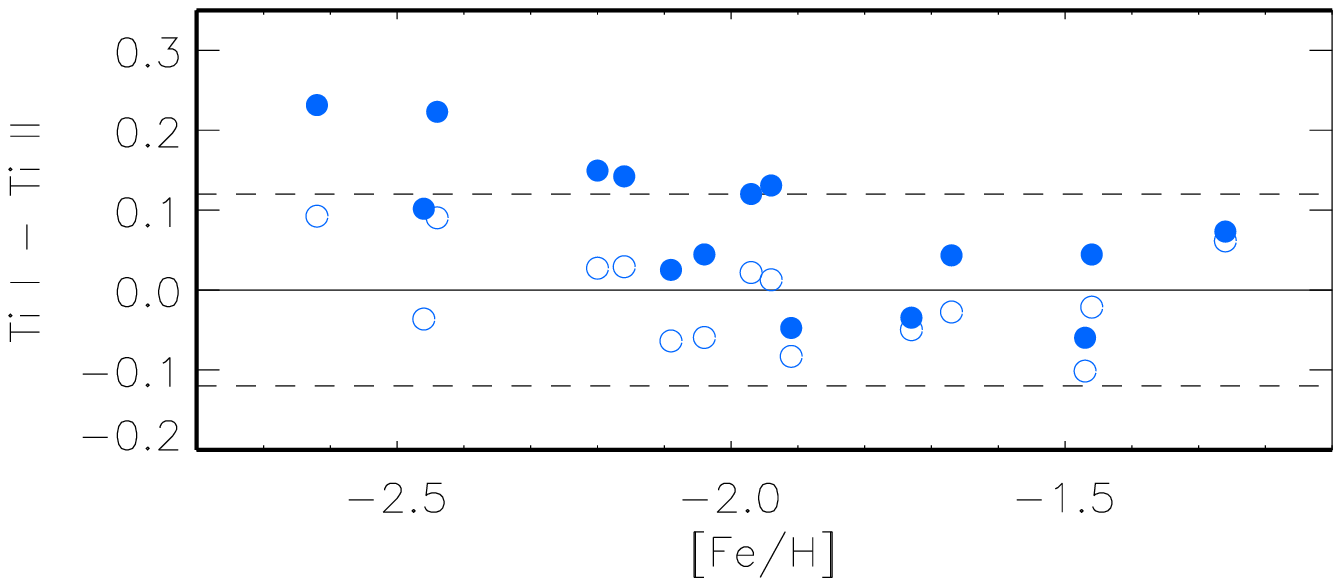}
	\includegraphics[width=80mm]{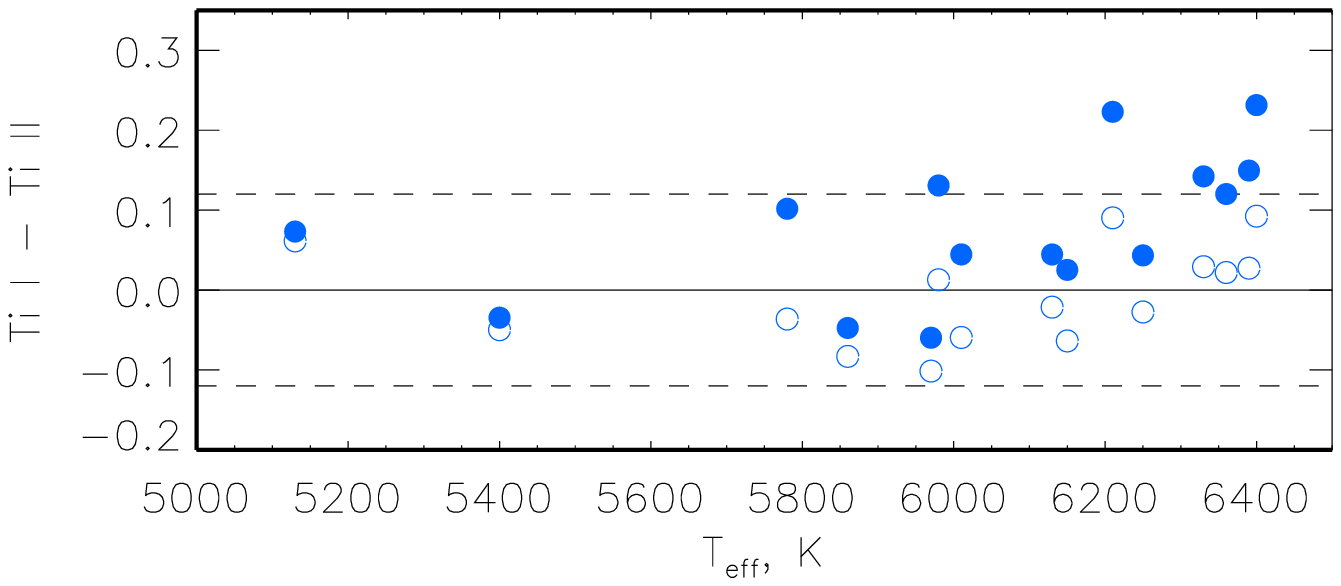}		
	\caption{The abundance difference Ti\ione--Ti\ii\ for the fifteen cool stars in LTE (open circles) and NLTE with \kH~=1  (filled circles). Dashed lines indicate a typical statistical error of $\rm \sqrt{ \sigma_{Ti I}+\sigma_{Ti II}}  = \pm$~0.12~dex.} 
	\label{ti12cool}
\end{figure*}

Metal-poor stars suit better for a calibration of \kH\ parameter than the solar-metallicity  stars. This is due to the deviations from LTE grow with decreasing [Fe/H] because of increasing the ultraviolet (UV) flux and decreasing electronic number density.
Our sample of cool MP dwarfs  includes 15  stars with  $-2.6 \leq$~[Fe/H]~$\leq -1.3$.
For  all the stars we determined the titanium abundance under the LTE assumption and in NLTE with \kH~=~1.0, and also with \kH~=~0.5 and 0.1 for few stars. The abundance differences Ti\ione--Ti\ii\ are listed in Table \ref{lines5} for various line formation scenarios and shown in Fig.~\ref{ti12cool} for LTE and NLTE with \kH~=~1.0.
For the seven stars the NLTE calculations result in consistent within the error bars abundances from Ti\ione\ and Ti\ii. 
For example, in HD~94028 Ti\ione--Ti\ii~=~$-0.11$~dex in LTE, and reduces to $-0.05$~dex in NLTE (\kH~=~1). 
For the other eight stars, on the contrary, an agreement between Ti\ione\ and Ti\ii\ is better in LTE compared to that in NLTE. Moreover, for these stars Ti\ione--Ti\ii~$\geq$~0 is obtained already in LTE, and the difference increases in NLTE.  For example, in BD~$-13^\circ3442$ we derived the largest discrepancy of 0.23~dex when using NLTE with \kH~=~1, while in LTE Ti\ione--Ti\ii~=~0.09~dex. 
All these stars, except HD~103095, are either turn-off (TO) stars  with 6200~$\le$~\teff~$\le$~6400~K, 3.9~$\le$~\lgg~$\le$~4.1,  $-2.6 \le$[Fe/H]$\le -1.9$, or VMP subgiants (SG) with \teff~$\ge$~5780~K. 
Due to lower \kH\ leads to larger NLTE effects, we do not perform calculations with \kH~$\le$~1 for these stars, except HD~84937. 
For the eight dwarfs with negative LTE abundance difference Ti\ione--Ti\ii\ we performed NLTE calculations with \kH~=~0.5.
The minimal difference Ti\ione--Ti\ii for maximal number of stars is achieved, when using \kH~=~1.

\underline{HD~122563 (MP giant)}. In LTE we derived an abundance difference of Ti\ione--Ti\ii$ = -0.36$~dex, and in NLTE it decreases in absolute value and amounts to Ti\ione--Ti\ii$ = -0.18$~dex, $-0.13$~dex, and $-0.06$~dex, when using \kH~=~1.0, 0.5, and 0.1, respectively. To achieve an agreement between Ti\ione\ and Ti\ii, the lower \kH\ is required, compared with that for the dwarfs. 
It is worth noting that similar conclusion was drawn by  \citet{mash_fe} from a relative to the Sun line-by-line differential analysis of iron lines in HD~122563. \citet{mash_fe} derived an abundance difference of Fe\ione--Fe\ii$ = -0.21$~dex in LTE and   Fe\ione--Fe\ii$ = -0.18$~dex, $-0.05$~dex, and 0.03~dex in NLTE, when using \kH~=~1.0, 0.1, and 0.0, respectively. While to achieve the Fe\ione/Fe\ii\ balance for  MP TO-star HD~84937  \kH~=~1 is required.
In HD~122563, for both Fe\ione\ and Fe\ii\ and Ti\ione\ and Ti\ii\ NLTE leads to smaller abundance difference between the two ionisation stages compared to LTE. 

\subsection{Comparison with other studies}
We have the four stars in common with \citet{Bergemann2011}, namely, the Sun, HD~84937, HD140283, and HD~122563.
For the common lines of Ti\ione\ and Ti\ii\ used in the solar analyses we recalculated abundances  derived by  \citet{Bergemann2011} using gf-values adopted in this study. For the majority lines the LTE abundance difference between \citet{Bergemann2011} and our data does not exceed 0.03~dex and nowhere exceeds 0.05~dex. We also compared the NLTE abundance corrections for Ti\ione. \citet{Bergemann2011} adopted \kH~=~3 in the NLTE calculations, while we use \kH~=~1. However, for the majority lines she computed larger NLTE abundance corrections, by up to 0.03~dex (for Ti\ione\ 4981\AA). \citet{Bergemann2011} derived with \kH~=~3 the average abundance difference Ti\ione$_{\rm NLTE}$--Ti\ione$_{\rm LTE}$~=~0.05~dex, while we obtain the same value, when using \kH~=~0.5.
The  smaller NLTE effects in this study compared with \citet{Bergemann2011} are due to using a comprehensive model atom that includes predicted high-excitation levels of Ti\ione.

The difference between our and \citet{Bergemann2011} NLTE results grows, when moving to the MP stars.
We compare the abundance differences Ti\ione$_{\rm NLTE}$--Ti\ione$_{\rm LTE}$ and Ti\ione\--Ti\ii.
For HD~84937,  \citet{Bergemann2011} derived Ti\ione$_{\rm NLTE}$--Ti\ione$_{ \rm LTE}$~=~0.14~dex using  \kH~=~3 and MAFAGS-OS model atmosphere \citep{Grupp2009}.
Using the same stellar parameters for this star, \kH~=~3, and MARCS model atmosphere \citep{MARCS} we derived Ti\ione$_{\rm NLTE}$--Ti\ione$_{ \rm LTE}$~=~0.09~dex.
We checked, whether this abundance discrepancy can be attributed to different codes for  model atmosphere calculation. 
We calculated Ti\ione\ and Ti\ii\ abundances with MARCS and MAFAGS-OS models, and found that the abundance difference does not exceed 0.02~dex for any line.
For HD~84937  \citet{Bergemann2011} derived in LTE Ti\ione--Ti\ii~=~0.11~dex, while we found Ti\ione--Ti\ii$_{ \rm LTE} = 0.03$~dex. 
For HD~140283 she presents abundances calculated only with the MAFAGS-ODF  model structure, Ti\ione--Ti\ii~=~0.02~dex in LTE and 0.16~dex  in NLTE (\kH~=~3). The corresponding values in our calculations are $-0.05$~dex (LTE) and 0.09~dex (NLTE, \kH~=~1).  
Similar situation in our studies was found for HD~122563. We derived discrepancies of  Ti\ione--Ti\ii$ = -0.36$~dex and $-0.18$~dex in  LTE and NLTE (\kH~=~1), respectively. 
The corresponding LTE and NLTE (\kH~=~3) values from \citet{Bergemann2011} are $-0.40$~dex and $-0.10$~dex.  
This abundance comparison indicates that our model atom leads to smaller deviations from LTE compared with those computed by \citet{Bergemann2011}.

The star HD~84937 is used like a reference star in many studies, since its atmospheric parameters are well-determined by different independent methods. 
\citet{Sneden2016} investigated the titanium lines under the LTE assumption adopting \teff~=~6300~K, log~g~=~4.0, [Fe/H]$ = -2.15$, \vt\ =~1.5~\kms\ and the interpolated model from \citet{Kurucz2011} model grid. In LTE they found consistent abundances from Ti\ione\ and Ti\ii. Using adopted in their study atmospheric parameters and our linelist we derived in LTE Ti\ione$-$Ti\ii~=~0.02~dex.
A very similar abundance difference of Ti\ione--Ti\ii~=~0.03~dex was found, with our parameters 6350/4.09/$-2.16$/1.7. 
This is due to higher \teff\ and higher \lgg\ lead to decrease in abundance from Ti\ione\ and Ti\ii, respectively, keeping the ionisation balance safe.
HD~84937 is one of the stars, where NLTE leads to positive Ti\ione--Ti\ii abundance difference, as discussed above.


\subsection{What is a source  of discrepancy between Ti\ione\ and Ti\ii\ in VMP TO-stars?} 
\underline{The treatment of collisions with H\ione.}
The main NLTE mechanism for Ti\ione\ is the UV overionisation and there is no process, which can result in strengthened lines of  Ti\ione\ and  negative NLTE abundance corrections.
Inelastic collisions with H\ione\ atoms serve as an additional source of thermalisation that reduces, but does not cancel the overionisation. It is worth noting that in the atmospheres of our VMP ([Fe/H]~$\leq -2$) TO-stars the lines of Ti\ione\ are weak (EW $\leq$~20~m\AA) and form inwards log$\tau_{5000}$~=~$-1$.  
The NLTE abundance corrections for Ti\ii\ lines are positive in the  model 6350/4.09/$-2.15$, $\Delta_{\rm NLTE} \leq$~0.01~dex when \kH~=~1, and $\Delta_{\rm NLTE}$ can be up to 0.08~dex, when neglecting collisions with H\ione\ atoms.
To what extent inelastic collisions with H\ione\ atoms can help to solve the problem of Ti\ione--Ti\ii\ in the MP TO stars remains unclear until accurate collisional data will be computed for both Ti\ione\ + H\ione\ and Ti\ii\ + H\ione.

\underline{Uncertainties in \teff.}
The lines of Ti\ione\ are more sensitive to \teff\ variation compared with Fe\ione\ lines, because of lower ionisation threshold for Ti\ione\ compared to Fe\ione. 
For example, we found an abundance shift of 0.09~dex for Ti\ione\ lines, and only 0.05~dex for  Fe\ione, when adopting 70~K lower \teff\  for HD~103095 (5130/4.66/$-1.26$). 
For this star, a downward revision of \teff\ by 70~K results in consistent abundances from Ti\ione\ and Ti\ii, and does not destroy Fe\ione/Fe\ii\ ionisation equilibrium.
However, a different situation was found for the VMP TO stars.
For example, we obtained similar abundance shifts of 0.08~dex and 0.06~dex for Ti\ione\ and Fe\ione, respectively, when adopting 100~K lower \teff\ for  HD~84937  (6350/4.09/$-2.16$). 
For this star, the \teff\ decrease results in the ionisation equilibrium for titanium, but not for iron.  

\begin{figure}
	\centering
	\includegraphics[width=80mm]{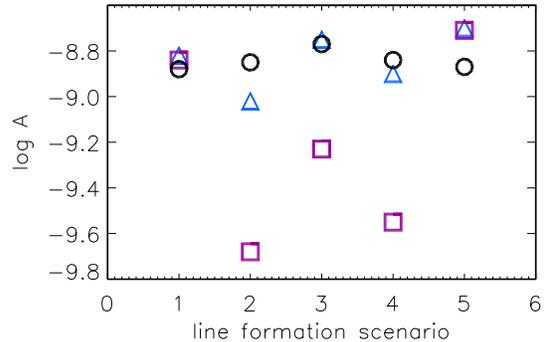}
	\caption{Titanium abundances in HD~84937 from individual lines:  Ti\ione\ 4617 \AA\ (triangle), Ti\ione\ 4681 \AA\ (square), Ti\ii\ 5336 \AA\ (circle) in different line formation scenarios, namely, 1~=~LTE+1D, 2~=~LTE+3D, 3~=~NLTE(\kH~=~0)+3D, 4~=~NLTE(\kH~=~1)+3D, 5~=~NLTE(\kH~=~1)+1D.} 
	\label{hd84937_3d1d}
\end{figure}

\underline{3D effects.} The solution of the NLTE problem with such a comprehensive model atom as treated in this study is only possible, at present, with classical plane-parallel (1D) model atmospheres. Neglecting atmospheric inhomogeneities (3D effects) can lead to errors in our results. From hydrodynamical modelling of stellar atmospheres \citet{Collet2007} and \citet{Dobrovolskas2013} predict negative  abundance corrections $\Delta_{\rm 3D}$~=~log~A$_{\rm 3D}$--log~A$_{\rm 1D}$ for lines of neutral species in red giant stars. 
In the models of TO (5900/4.0) stars with  [Fe/H]$ = -2$, $\Delta_{\rm 3D}$ increases in absolute value with decreasing the excitation energy of the lower level, and reaches $-0.84$~dex and $-0.20$~dex for the $\lambda$~=~4000~\AA\ lines with \eexc~=~0 and 2~eV, respectively  \citep{Dobrovolskas_phd}.
All the lines of Ti\ione\ used for our MP TO stars have \eexc~$\le $1.75~eV.
The 3D abundance corrections can be either positive or negative, and do not exceed 0.07~dex in absolute value for the lines of Ti\ii. 
Negative  3D corrections for Ti\ione\ could help to achieve an agreement between Ti\ione\ and Ti\ii.
We selected two lines of Ti\ione, at 4617~\AA\ (\eexc~=~1.75~eV) and 4681~\AA\ (\eexc~=~0.05~eV), and Ti\ii\ 5336~\AA\ (\eexc~=~1.58~eV), which give consistent within 0.02~dex LTE abundances  and calculated the abundance differences  Ti\ione--Ti\ii for different line formation scenarios, taking 3D abundance corrections from  \citet{Dobrovolskas_phd}. Abundances from individual lines are shown in Fig.~\ref{hd84937_3d1d}.
In LTE we derived Ti\ione--Ti\ii~=~0.05~dex and $-0.49$~dex in 1D and 3D, respectively. 
In NLTE+3D we derived $-0.27$~dex (\kH~=~1) and $-0.16$~dex (\kH~=~0), while  Ti\ione--Ti\ii=~0.17~dex in NLTE(\kH~=~1)+1D, which is our standard scenario. 
The predicted 3D effects are too strong for low-excitation lines of Ti\ione\ and produce a large discrepancy between Ti\ione\ lines with different \eexc, which reaches 0.66~dex in LTE+3D.
We suppose that for MP stars simple co-adding the NLTE(1D) and 3D(LTE) corrections is too rough procedure, because both NLTE and 3D effects are equally significant. 

\underline{Chromospheres.} One more source can be connected with a star's chromosphere that heats the line formation layers. An inspiring insight into this problem was presented by \citet{Dupree2016}. Further efforts should be invested to evaluate a possible influence of the star's chromosphere on the formation of titanium lines.

\begin{table}
	\caption{The abundance difference Ti\ione\--Ti\ii\ for cool stars of the sample in different line formation scenarios.}
	\renewcommand\arraystretch{1.1}
	\vspace{2mm}
	\begin{large}
		\centering
		\vspace{5mm}
		\label{lines5}

		\begin{tabular}{|l|r|r|r|r|}
                          &        & \multicolumn{3}{c}{NLTE, \kH} \\
                          &        &  \multicolumn{3}{c}{-----------------------------}  \\  
                          Star                    &  LTE   &   1 &  0.5 &  0.1 \\
			\hline
Sun  & --0.05 & --0.03 & --0.02 &  0.00 \\
HD~64090 & --0.04 & --0.01 &   0.00 &  0.06 \\
HD~84937 &   0.03 &   0.15 &   0.19 &  0.24 \\ 
HD~94028 & --0.11 & --0.05 & --0.03 &  0.05 \\ 
HD~122563 & --0.36 & --0.18 & --0.13 & --0.06 \\
BD+07$^\circ$~4841 & --0.06 & 0.02  & 0.06 &  \\ 
BD+09$^\circ$~0352 & --0.06 & 0.04 & 0.08  &  \\ 
HD~140283 & --0.05 &   0.09 &   0.14 &       \\ 
BD+29$^\circ$~2091 & --0.14 & --0.09 & --0.07  &  \\ 
G~090-003   & --0.07 & 0.05  & 0.09  &  \\
HD~24289 &  0.00 & 0.14 & & \\ 
HD~74000 &   0.01 &   0.12 &        &       \\ 
HD~103095 &  0.06  & 0.07 & & \\
HD~108177 & --0.07 & 0.02 &  &  \\ 
BD--13$^\circ$~3442 &   0.09 &   0.23 &      &  \\ 
BD--04$^\circ$~3208 &   0.02 &   0.15 &      &  \\ 
BD+24$^\circ$~1676 &   0.07 & 0.20 &         &  \\ 
			\hline
		\end{tabular}
	\end{large}
\end{table}

\begin{table*}
	\caption{Average NLTE and LTE abundances from Ti\ione\ and Ti\ii\ lines in the programme stars.}
	\renewcommand\arraystretch{1.1}
	\vspace{2mm}
	\begin{large}
		\centering
		\vspace{5mm}
		\label{abund}

		\begin{tabular}{|l|c|c|c|c|c|c|}
					\hline
			Star   &  N$_{\rm Ti I}$   & log A(Ti\ione)$_{\rm LTE}$ & log A(Ti\ione)$_{\rm NLTE}$ &  N$_{\rm Ti II}$   & log A(Ti\ii)$_{\rm LTE}$ & log A(Ti\ii)$_{\rm NLTE}$ \\
			\hline
		HD~37594 &   8 &  --7.23$\pm$0.13 &  --7.11$\pm$0.11 &  27 &  --7.01$\pm$0.15 &  --7.04$\pm$0.11 \\
		HD~32115 &   6 &  --7.45$\pm$0.05 &  --7.31$\pm$0.05 &   9 &  --7.23$\pm$0.07 &  --7.26$\pm$0.05 \\
		HD~72660 &   5 &  --6.63$\pm$0.05 &  --6.57$\pm$0.08 &  36 &  --6.54$\pm$0.12 &  --6.59$\pm$0.08 \\
		HD~73666 &   2 &  --6.94$\pm$0.02 &  --6.89$\pm$0.09 &   6 &  --6.72$\pm$0.20 &  --6.84$\pm$0.09 \\
		HD~145788 & & & &  32 &  --6.76$\pm$0.15 &  --6.81$\pm$0.07   \\
		Sirius &  & & & 6 &  --6.84$\pm$0.06 &  --6.89$\pm$0.04   \\
		21~Peg  & & & &  46 &  --7.24$\pm$0.05 &  --7.24$\pm$0.04   \\
		$\pi$~Cet & & & &  11 &  --7.41$\pm$0.09 &  --7.14$\pm$0.08  \\
		\hline
		Sun &  27 &  --7.11$\pm$0.05 &  --7.09$\pm$0.05 &  12 &  --7.06$\pm$0.04 &  --7.06$\pm$0.04 \\
		BD--13$^\circ$ 3442 &   3 &  -9.25$\pm$0.04 &  -9.09$\pm$0.04 &  15 &  -9.34$\pm$0.06 &  -9.32$\pm$0.06 \\		
		BD--04$^\circ$ 3208 &   9 &  -8.90$\pm$0.05 &  -8.77$\pm$0.05 &  17 &  -8.92$\pm$0.06 &  -8.92$\pm$0.05 \\
		BD+7$^\circ$ 4841 &  26 &  --8.24$\pm$0.05 &  --8.17$\pm$0.05 &  34 &  --8.17$\pm$0.06 &  --8.19$\pm$0.05 \\
		BD+9$^\circ$ 0352 &   9 &  --8.87$\pm$0.05 &  --8.78$\pm$0.05 &  22 &  --8.81$\pm$0.05 &  --8.82$\pm$0.04 \\
		BD+24$^\circ$ 1676 &   7 &  -9.12$\pm$0.06 &  -8.98$\pm$0.06 &  16 &  -9.19$\pm$0.06 &  -9.18$\pm$0.06 \\
		BD+29$^\circ$ 2091 &  20 &  --8.76$\pm$0.06 &  --8.72$\pm$0.06 &  24 &  --8.62$\pm$0.08 &  --8.63$\pm$0.07 \\
		HD~24289 &  16 &  --8.79$\pm$0.10 &  --8.67$\pm$0.10 &  27 &  --8.79$\pm$0.08 &  --8.81$\pm$0.09 \\
		HD~64090 &  35 &  --8.73$\pm$0.07 &  --8.71$\pm$0.07 &  30 &  --8.69$\pm$0.06 &  --8.70$\pm$0.05 \\
		HD~74000 &  15 &  -8.78$\pm$0.07 &  -8.68$\pm$0.07 &  26 &  -8.79$\pm$0.08 &  -8.80$\pm$0.08 \\
		HD~84937 &  12 &  --8.84$\pm$0.04 &  --8.71$\pm$0.04 &  15 &  --8.87$\pm$0.08 &  --8.86$\pm$0.08 \\
		HD~94028 &  26 &  --8.34$\pm$0.06 &  --8.30$\pm$0.07 &  26 &  --8.23$\pm$0.04 &  --8.24$\pm$0.05 \\
		HD~103095 &  37 &  --8.06$\pm$0.09 &  --8.05$\pm$0.09 &  29 &  --8.12$\pm$0.07 &  --8.12$\pm$0.07 \\
		HD~108177 &  14 &  --8.50$\pm$0.06 &  --8.43$\pm$0.07 &  12 &  --8.43$\pm$0.07 &  --8.45$\pm$0.06 \\
		HD~122563 &  22 &  --9.82$\pm$0.07 &  --9.64$\pm$0.08 &  36 &  --9.46$\pm$0.06 &  --9.46$\pm$0.07 \\
		HD~140283 &  19 &  --9.36$\pm$0.07 &  --9.21$\pm$0.07 &  25 &  --9.31$\pm$0.05 &  --9.30$\pm$0.05 \\
		G~090--03 &  18 &  --8.85$\pm$0.07 &  --8.75$\pm$0.07 &  30 &  --8.78$\pm$0.07 &  --8.79$\pm$0.06 \\
			\hline		
\multicolumn{7}{l}{For the cool stars the NLTE abundances were derived using \kH~=~1.} \\
			\hline
		\end{tabular}
	\end{large}
\end{table*}

\section{Conclusions}
\label{con}

We construct a comprehensive model atom for Ti\ione--\ii\ using  the energy levels from laboratory measurements and theoretical predictions and quantum mechanical photoionisation cross-sections.
NLTE line formation for Ti\ione\ and Ti\ii\ lines was considered in 1D-LTE model atmospheres of the 25 reference stars with reliable stellar parameters, which cover a broad range of effective temperatures 4600~$\le$~\teff~$\le$~12800~K, surface gravities 1.60~$\le$~\lgg~$\le$~4.70, and metallicities $-2.5~\le$~[Fe/H]~$\le$~+0.4.

The NLTE calculations for Ti\ione--\ii\ in A-type stars were performed for the first time. The NLTE titanium abundances were determined for the eight stars. For the four stars with both Ti\ione\ and Ti\ii\ lines observed, NLTE analysis provides consistent within 0.07~dex abundances from Ti\ione\ and Ti\ii\ lines, while the corresponding LTE abundance difference can be up to 0.22~dex in absolute value. For each species, NLTE leads to smaller line-to-line scatter compared with LTE. 
For stars with \teff $\ge$~7000~K  lines of Ti\ione\ and Ti\ii\ can be used for atmospheric parameter determination, when taking into account deviations from LTE. For the 22 lines of Ti\ione\ and 82 lines of Ti\ii\ we calculated the NLTE abundance corrections in a grid of model atmospheres with \teff\ from 6500~K to 13000~K, \lgg~=~4, [Fe/H]~=~0, and \vt~=2~\kms. 

We made progress in  determination of NLTE abundance of titanium for cool stars compared with data from the literature.
Taking into account a bulk of the predicted high-excitation levels of Ti\ione\ in the model atom  established close collisional coupling of the Ti\ione\ levels near the continuum to the ground state of Ti\ii\ resulting in smaller NLTE effects in cool model atmospheres compered with the \citet{Bergemann2011} data.
Because no accurate calculations of inelastic collisions of titanium with neutral hydrogen atoms are available, we use the Drawinian formalism with the scaling factor, which was estimated as \kH~=~1 from abundance comparison between Ti\ione\ and Ti\ii\ in the sample of cool main sequence stars over wide metallicity range, $-2.6 \leq$~[Fe/H]~$\leq$~0.0.  
For the VMP TO-stars NLTE fails to achieve agreement between  Ti\ione\ and Ti\ii. Moreover, for these stars  we derived positive abundance difference Ti\ione--Ti\ii in LTE, and it increases in NLTE. 
To clarify this matter, accurate collisional data for Ti\ione\ and Ti\ii\ would be extremely helpful. 

\section{Appendix}\label{apend}

In this section we present Table~\ref{atomic} and Table~\ref{corrections} in theirs entirety.

This is a full version of Table~\ref{atomic}. The list of Ti\ione\ and Ti\ii\ lines with the adopted atomic data. \\
		$\lambda$ (\AA), \eexc (eV),  log~gf,  transition,  log~$\gamma_{rad}$, log~$\gamma_4/N_e$, log~$\gamma_6/N_H $ \\

Ti I \\ 
4008.927   0.021  -1.000    3a3F --   y3F       8.000      -6.080      -7.750  \\
4060.262   1.052  -0.690    a3P  --   x3P       8.050      -6.050      -7.646   \\                     
4060.262   1.052  -0.690    a3P  --   x3P       8.050      -6.050      -7.646     \\                     
4287.403   0.836  -0.370    a5F  --   x5D       8.230      -6.010      -7.570    \\                       
4449.143   1.886   0.470    a3G  --   v3G       8.120      -5.560      -7.579    \\                           
4453.699   1.872   0.100    a3G  --   v3G       8.110      -4.970      -7.582    \\                       
4512.733   0.836  -0.400    a5F  --   y5F       8.130      -5.120      -7.593    \\                       
4533.240   0.848   0.540    a5F  --   y5F       8.130      -5.120      -7.593     \\                   
4534.776   0.836   0.350    a5F  --   y5F       8.130      -5.280      -7.596    \\  
4548.763   0.826  -0.280    a5F  --   y5F       8.130      -5.410      -7.598     \\     
4555.484   0.848  -0.400    a5F  --   y5F       8.130      -5.280      -7.596    \\                
4617.268   1.748   0.440    a5P  --   w5D       8.080      -5.860      -7.626     \\          
4623.097   1.739   0.160    a5P  --   w5D       8.070      -5.850      -7.627     \\               
4639.361   1.739  -0.050    a5P  --   w5D       8.070      -5.840      -7.740      \\     
4639.661   1.748  -0.140    a5P  --   w5D       8.070      -5.850      -7.740        \\                       
4639.940   1.733  -0.160    a5P  --   w5D       8.070      -5.840      -7.630     \\   
4656.468   0.000  -1.290$^2$    2a3F --   z3G       6.380      -6.110      -7.706     \\         
4681.909   0.050  -1.030$^2$    4a3F --   z3G       6.460      -6.110      -7.702     \\                               
4758.118   2.248   0.510    a3H  --   x3H       8.080      -6.040      -7.621             \\                 
4759.269   2.255   0.590    a3H  --   x3H       8.080      -6.040      -7.620      \\                
4820.410   1.502  -0.380    a1G  --   y1F       8.210      -5.940      -7.625     \\                         
4840.874   0.899  -0.430    a1D  --   y1D       7.530      -6.120      -7.697    \\               
4913.615   1.872   0.220    a3G  --   y3H       7.850      -5.890      -7.619     \\    
4981.731   0.848   0.570    a5F  --   y5G       7.950      -6.050      -7.626      \\
4991.067   0.836   0.450    a5F  --   y5G       7.940      -6.050      -7.629      \\             
4997.093   0.000  -2.070    2a3F --   z3D       6.900      -6.110      -7.722      \\         
4999.502   0.826   0.320    a5F  --   y5G       7.940      -6.050      -7.632      \\   
5009.645   0.021  -2.200    3a3F --   z3D       6.870      -6.110      -7.720      \\                        
5016.161   0.848  -0.480    a5F  --   y5G       7.940      -6.050      -7.629      \\    
5020.025   0.836  -0.330    a5F  --   y5G       7.940      -6.050      -7.630      \\                   
5024.844   0.818  -0.530    a5F  --   y5G       7.940      -6.050      -7.635      \\                
5025.570   2.041   0.250$^1$    z5G  --   e5F       7.960      -5.310      -7.550    \\                            
5036.464   1.443   0.140    b3F  --   w3G       8.160      -5.700      -7.539    \\                      
5039.955   0.021  -1.080    3a3F --   z3D       6.900      -6.110      -7.720    \\           
5064.652   0.048  -0.940    4a3F --   z3D       6.870      -6.110      -7.719    \\           
5147.477   0.000  -1.940    2a3F --   z3F       6.820      -6.110      -7.727    \\                          
5173.740   0.000  -1.060    2a3F --   z3F       6.820      -6.110      -7.729    \\     
5192.969   0.021  -0.950    3a3F --   z3F       6.820      -6.110      -7.727    \\      
5210.384   0.048  -0.820    4a3F --   z3F       6.810      -6.110      -7.724     \\   
5512.524   1.460  -0.400    b3F  --   w3D       8.190      -6.100      -7.700    \\                          
5514.343   1.429  -0.660    b3F  --   w3D       8.140      -5.990      -7.710    \\                          
5514.532   1.443  -0.500    b3F  --   w3D       8.150      -6.070      -7.710    \\                          
5866.449   1.066  -0.790    a3P  --   y3D       8.000      -6.070      -7.724    \\                  
6258.099   1.443  -0.390    b3F  --   y3G       8.250      -5.990      -7.582    \\      
6261.096   1.429  -0.530    b3F  --   y3G       8.260      -5.980      -7.585     \\          
8426.506   0.826  -1.200$^2$    a5F  --   z5D       6.370      -6.090      -7.711    \\
Ti II \\ 
2827.114   3.687  -0.020$^4$    z4G  --   e4G   8.850  -5.850  -7.720                                               \\ 
2828.077   3.749   0.870$^3$    z4G  --   e4H   8.860  -5.820  -7.720                                               \\ 
2834.011   3.716   0.000$^4$    z4G  --   e4G   8.850  -5.850  -7.720                                               \\ 
2841.935   0.607  -0.590    a2F  --   y2F   8.420  -6.390  -7.830                                         \\ 
2851.101   1.221  -0.730    a2P  --   x2D   8.320  -6.470  -7.820                                         \\ 
2853.931   0.607  -1.550    a2F  --   y2F   8.350  -6.390  -7.840                                         \\ 
2868.741   0.574  -1.380    a2F  --   y2D   8.260  -6.390  -7.850                                         \\ 
4012.385   0.574  -1.780    a2F  --   z4G   8.220  -6.390  -7.860                                        \\ 
4028.343   1.891  -0.920    b2G  --   y2F   8.420  -6.410  -7.830                                        \\ 
4053.820   1.892  -1.070    b2G  --   y2F   8.350  -6.410  -7.840                                  \\ 
4161.530   1.084  -2.090    a2D  --   z4D   8.410  -6.430  -7.840                                           \\ 
4163.640   2.589  -0.130    b2F  --   x2D   8.320  -6.470  -7.820                                           \\ 
4174.070   2.598  -1.260$^4$    b2F  --   x2D   8.320  -6.470  -7.820                                                 \\ 
4188.987   5.423  -0.600$^4$    y2G  --   e2G   8.900  -5.690  -7.690                                               \\ 
4190.233   1.084  -3.122$^1$    a2D  --   z4D   8.410  -6.430  -7.840                                               \\ 
4287.870   1.080  -1.790$^4$    a2D  --   z2D   8.170  -6.430  -7.850                                                 \\ 
4290.215   1.164  -0.870    a4P  --   z4D   8.410  -6.500  -7.840                                        \\ 
4300.049   1.180  -0.460    a4P  --   z4D   8.410  -6.490  -7.840                                        \\ 
4301.920   1.160  -1.210    a4P  --   z4D   8.410  -6.490  -7.840                                           \\ 
4316.794   2.047  -1.620    b2P  --   z2P   8.420  -6.460  -7.840                                        \\ 
4337.915   1.080  -0.960$^4$    a2D  --   z2D   8.160  -6.430  -7.850                                        \\ 
4374.820   2.060  -1.570    b2P  --   y2D   8.260  -6.460  -7.850                                           \\ 
4386.844   2.598  -0.960$^4$    b2F  --   y2G   8.450  -6.540  -7.830                                          \\ 
4391.020   1.231  -2.300    b4P  --   z4D   8.410  -6.410  -7.840                                           \\ 
4394.059   1.221  -1.770    a2P  --   z4D   8.410  -6.490  -7.840                  \\ 
4395.031   1.084  -0.540    a2D  --   z2F   8.160  -6.430  -7.850                                  \\ 
4395.839   1.242  -1.930    b4P  --   z4D   8.410  -6.410  -7.840                 \\ 
4399.772   1.236  -1.200    a2P  --   z4D   8.410  -6.500  -7.840                  \\ 
4409.235   1.242  -2.780    b4P  --   z4D   8.410  -6.410  -7.840                                         \\ 
4409.520   1.231  -2.530    b4P  --   z4D   8.410  -6.410  -7.840                                \\ 
4411.070   3.093  -0.650    c2D  --   x2F   8.290  -6.330  -7.840                                 \\ 
4411.925   1.224  -2.620    b4P  --   z4D   8.420  -6.410  -7.840                \\ 
4417.713   1.165  -1.190$^4$    a4P  --   z2D   8.170  -6.580  -7.850           \\ 
4418.331   1.236  -1.990    a2P  --   z4D   8.410  -6.490  -7.840     \\ 
4421.938   2.060  -1.640    b2P  --   z2P   8.350  -6.460  -7.840                                   \\ 
4423.239   1.231  -3.066$^1$    b4P  --   z4D   8.420  -6.410  -7.840                                               \\ 
4432.109   1.236  -3.080    a2P  --   z4D   8.420  -6.490  -7.840                                         \\ 
4441.730   1.180  -2.330$^4$    a4P  --   z2D   8.170  -6.580  -7.850                                       \\ 
4443.801   1.080  -0.710    a2D  --   z2F   8.150  -6.430  -7.850             \\ 
4444.554   1.115  -2.200    a2G  --   z2F   8.160  -6.590  -7.850                              \\ 
4450.482   1.084  -1.520    a2D  --   z2F   8.150  -6.430  -7.850            \\ 
4464.449   1.161  -1.810$^4$    a4P  --   z2D   8.160  -6.600  -7.850                  \\ 
4468.500   1.130  -0.630    a2G  --   z2F   8.860  -5.710  -7.690                                       \\ 
4468.510   1.130  -0.630    a2G  --   z2F   8.860  -5.710  -7.690                                   \\ 
4469.151   1.084  -2.550    a2D  --   z4F   8.380  -6.430  -7.840                                 \\ 
4470.853   1.165  -2.020$^4$    a4P  --   z2D   8.160  -6.600  -7.850                  \\ 
4488.324   3.122  -0.500    c2D  --   x2F   8.280  -6.330  -7.840                                           \\ 
4501.270   1.115  -0.770    a2G  --   z2F   8.150  -6.590  -7.850                          \\ 
4518.330   1.080  -2.560    a2D  --   z4F   8.380  -6.430  -7.840                                           \\ 
4529.474   1.571  -1.750    a2H  --   z2G   8.310  -6.490  -7.820                                         \\ 
4533.960   1.237  -0.530$^4$    a2P  --   z2D   8.170  -6.540  -7.850                                     \\ 
4544.020   1.243  -2.580$^4$    a2G  --   z4F   8.170  -6.410  -7.850                                \\ 
4549.620   1.583  -0.220    a2H  --   z2G   8.310  -6.490  -7.820                                           \\ 
4563.757   1.221  -0.795$^1$    a2P  --   z2D   8.160  -6.550  -7.850                        \\ 
4568.314   1.224  -3.030$^5$    b4P  --   z2D   8.160  -6.410  -7.850                               \\ 
4571.971   1.571  -0.310    a2H  --   z2G   8.310  -6.490  -7.820                             \\ 
4583.410   1.164  -2.840    a4P  --   z2F   8.150  -6.590  -7.850                         \\ 
4589.958   1.237  -1.620$^5$    a2P  --   z2D   8.160  -6.540  -7.850                                              \\ 
4636.320   1.165  -3.024$^1$    a4P  --   z4F   8.380  -6.500  -7.840                                            \\ 
4657.201   1.242  -2.290    b4P  --   z2F   8.160  -6.410  -7.850                         \\ 
4708.663   1.236  -2.350    a2P  --   z2F   8.150  -6.540  -7.850                        \\ 
4719.515   1.242  -3.320    b4P  --   z2F   8.150  -6.410  -7.850                                         \\ 
4763.880   1.221  -2.400    a2P  --   z4F   8.380  -6.510  -7.840                                           \\ 
4764.525   1.236  -2.690    a2P  --   z4F   8.380  -6.500  -7.840                                \\ 
4779.985   2.048  -1.260$^5$    b2P  --   z2S   8.230  -6.460  -7.860                                         \\ 
4798.530   1.080  -2.660    a2D  --   z4G   8.220  -6.430  -7.860                          \\ 
4805.085   2.061  -0.960$^5$    b2P  --   z2S   8.230  -6.460  -7.860                                   \\ 
4865.612   1.115  -2.700    a2G  --   z4G   8.220  -6.490  -7.860                                  \\ 
4911.190   3.122  -0.640    c2D  --   y2P   8.260  -6.330  -7.830                                  \\ 
4996.367   1.582  -3.290$^6$    b2D2 --   z4D   8.410  -6.490  -7.840                                               \\ 
5005.157   1.565  -2.730    b2D2 --   z4D   8.410  -6.490  -7.840                                \\ 
5010.210   3.093  -1.350    c2D  --   x2D   8.320  -6.330  -7.820                                           \\ 
5013.330   3.095  -2.028$^1$    c2D  --   x2D   8.320  -6.330  -7.820                                               \\ 
5013.686   1.581  -2.140    b2D2 --   z4D   8.410  -6.500  -7.840                                       \\ 
5072.290   3.122  -1.020    c2D  --   x2D   8.320  -6.330  -7.820                                           \\ 
5129.160   1.891  -1.340    b2G  --   z2G   8.310  -6.410  -7.820                          \\ 
5154.070   1.566  -1.750$^4$    b2D2 --   z2D   8.170  -6.580  -7.850                                \\ 
5185.913   1.892  -1.410    b2G  --   z2G   8.310  -6.410  -7.820                  \\ 
5188.680   1.582  -1.050$^4$    b2D2 --   z2D   8.170  -6.580  -7.850                                              \\ 
5211.536   2.589  -1.410    b2F  --   y2F   8.420  -6.480  -7.830                                    \\ 
5226.550   1.570  -1.260$^4$    b2D2 --   z2D   8.160  -6.590  -7.850                                        \\ 
5262.140   1.582  -2.250$^4$    b2D2 --   z2D   8.160  -6.590  -7.850                                                 \\ 
5268.610   2.597  -1.610    b2F  --   y2F   8.350  -6.480  -7.840                                           \\ 
5336.786   1.581  -1.600    b2D2 --   z2F   8.160  -6.590  -7.850     \\ 
5381.022   1.565  -1.970    b2D2 --   z2F   8.150  -6.590  -7.850                 \\ 
5418.768   1.581  -2.130    b2D2 --   z2F   8.150  -6.590  -7.850                 \\ 
5490.690   1.566  -2.663$^1$    b2D2 --   z4F   8.380  -6.510  -7.840                                       \\ 
6491.566   2.061  -1.942$^1$    b2P  --   z2D   8.170  -6.460  -7.850                                              \\ 
6606.950   2.060  -2.790$^3$    b2P  --   z2D   8.160  -6.460  -7.850                                        \\ 
6680.133   3.093  -1.890    c2D  --   y2F   8.350  -6.330  -7.840                                    \\ 
6998.905   3.122  -1.280    c2D  --   y2D   8.260  -6.330  -7.850            \\ 
sources of gf-values \\
1 - Kurucz, \\
2 - BLNP, Blackwell-Whitehead, R.~J. and Lundberg, H. and Nave, G. and Pickering, J.~C. and Jones, H.~R.~A. and Lyubchik, Y. and Pavlenko, Y.~V. and Viti, S., Monthly Notices Roy. Astron. Soc., 373, 1603-1609 (2006); \\
3 - MFW, Martin, G.A. and Fuhr, J.R. and Wiese, W.L., J. Phys. Chem. Ref. Data Suppl., 17, 3 (1988); \\
4 - PTP, Pickering, J.~C. and Thorne, A.~P. and Perez, R., Astrophys. J. Suppl. Ser., 132, 403-409 (2001); \\
5 - RHL, Ryabchikova, T.~A. and Hill, G.~M. and Landstreet, J.~D. and Piskunov, N. and Sigut, T.~A.~A., Monthly Notices Roy. Astron. Soc., 267, 697 (1994);  \\
6 - BHN, Bizzarri, A. and Huber, M.~C.~E. and Noels, A. and Grevesse, N. and Bergeson, S.~D. and Tsekeris, P. and Lawler, J.~E., Astronomy and Astrophysics, 273, 707 (1993);  \\
gf-values taken from Wisconsin  \citep{Lawler2013_ti1,Wood2013_ti2} if not prescribed.  \\

This is a full version of Table~\ref{corrections}. NLTE abundance corrections and equivalent widths for the lines of Ti\ione\ and Ti\ii\ depending on \teff\ in the models with \lgg~=~4, [Fe/H]~=~0, and \vt~=~2~\kms. A portion is shown here for guidance regarding its form and content. If EW~=~$-1$ and \dnlte~=~$-1$, this means that EW~$<$~5~m\AA\ in a given model atmosphere.

Calculations are performed for the 27 following effective temperatures (K): \\
6500  6750  7000  7250  7500  7750  8000  8250  8500  8750  9000  9250  9500  9750 10000 10250 10500 10750 11000 11250 11500 11750 12000 12250 12500 12750 13000 \\

The file is constructed as following: \\
wavelength, A; Ti species; excitation energy, eV; gf-value \\
equivalent width$_1$; ...; equivalent width$_{27}$ \\
NLTE abundance correction$_1$; ...; NLTE abundance correction$_{27}$ \\

 4287.4028  A  Ti1  Eexc =  0.836  log gf =    -0.370 \\ 
 38    29    21    15    11     7     5    -1    -1    -1    -1    -1    -1    -1    -1    -1    -1    -1    -1    -1    -1    -1    -1    -1    -1    -1    -1 \\ 
 0.11  0.10  0.09  0.08  0.08  0.08  0.08 -1.00 -1.00 -1.00 -1.00 -1.00 -1.00 -1.00 -1.00 -1.00 -1.00 -1.00 -1.00 -1.00 -1.00 -1.00 -1.00 -1.00 -1.00 -1.00 -1.00 \\ 
 4453.6992  A  Ti1  Eexc =  1.872  log gf =     0.100 \\ 
 18    13    10     7     5    -1    -1    -1    -1    -1    -1    -1    -1    -1    -1    -1    -1    -1    -1    -1    -1    -1    -1    -1    -1    -1    -1 \\ 
 0.14  0.13  0.12  0.12  0.12 -1.00 -1.00 -1.00 -1.00 -1.00 -1.00 -1.00 -1.00 -1.00 -1.00 -1.00 -1.00 -1.00 -1.00 -1.00 -1.00 -1.00 -1.00 -1.00 -1.00 -1.00 -1.00 \\ 
 4512.7329  A  Ti1  Eexc =  0.836  log gf =    -0.400 \\ 
 38    28    21    15    10     7     5    -1    -1    -1    -1    -1    -1    -1    -1    -1    -1    -1    -1    -1    -1    -1    -1    -1    -1    -1    -1 \\ 
 0.10  0.09  0.08  0.08  0.08  0.08  0.09 -1.00 -1.00 -1.00 -1.00 -1.00 -1.00 -1.00 -1.00 -1.00 -1.00 -1.00 -1.00 -1.00 -1.00 -1.00 -1.00 -1.00 -1.00 -1.00 -1.00 \\ 
 4533.2402  A  Ti1  Eexc =  0.848  log gf =     0.540 \\ 
 94    82    72    60    49    37    27    18    12     7    -1    -1    -1    -1    -1    -1    -1    -1    -1    -1    -1    -1    -1    -1    -1    -1    -1 \\ 
 0.04  0.05  0.05  0.06  0.07  0.08  0.09  0.10  0.11  0.11 -1.00 -1.00 -1.00 -1.00 -1.00 -1.00 -1.00 -1.00 -1.00 -1.00 -1.00 -1.00 -1.00 -1.00 -1.00 -1.00 -1.00 \\ 
 4534.7759  A  Ti1  Eexc =  0.836  log gf =     0.350 \\ 
 86    74    63    52    41    30    21    14     9     5    -1    -1    -1    -1    -1    -1    -1    -1    -1    -1    -1    -1    -1    -1    -1    -1    -1 \\ 
 0.06  0.06  0.06  0.07  0.07  0.08  0.09  0.10  0.11  0.11 -1.00 -1.00 -1.00 -1.00 -1.00 -1.00 -1.00 -1.00 -1.00 -1.00 -1.00 -1.00 -1.00 -1.00 -1.00 -1.00 -1.00 \\ 
 4548.7632  A  Ti1  Eexc =  0.826  log gf =    -0.280 \\ 
 46    35    26    19    13     9     6    -1    -1    -1    -1    -1    -1    -1    -1    -1    -1    -1    -1    -1    -1    -1    -1    -1    -1    -1    -1 \\ 
 0.10  0.09  0.08  0.08  0.08  0.08  0.09 -1.00 -1.00 -1.00 -1.00 -1.00 -1.00 -1.00 -1.00 -1.00 -1.00 -1.00 -1.00 -1.00 -1.00 -1.00 -1.00 -1.00 -1.00 -1.00 -1.00 \\ 
 4617.2681  A  Ti1  Eexc =  1.748  log gf =     0.440 \\ 
 40    31    24    18    13     9     7    -1    -1    -1    -1    -1    -1    -1    -1    -1    -1    -1    -1    -1    -1    -1    -1    -1    -1    -1    -1 \\ 
 0.13  0.13  0.13  0.13  0.12  0.13  0.12 -1.00 -1.00 -1.00 -1.00 -1.00 -1.00 -1.00 -1.00 -1.00 -1.00 -1.00 -1.00 -1.00 -1.00 -1.00 -1.00 -1.00 -1.00 -1.00 -1.00 \\ 
 4656.4678  A  Ti1  Eexc =  0.000  log gf =    -1.290 \\ 
 25    17    12     8     5    -1    -1    -1    -1    -1    -1    -1    -1    -1    -1    -1    -1    -1    -1    -1    -1    -1    -1    -1    -1    -1    -1 \\ 
 0.20  0.19  0.17  0.16  0.15 -1.00 -1.00 -1.00 -1.00 -1.00 -1.00 -1.00 -1.00 -1.00 -1.00 -1.00 -1.00 -1.00 -1.00 -1.00 -1.00 -1.00 -1.00 -1.00 -1.00 -1.00 -1.00 \\ 
 4759.2690  A  Ti1  Eexc =  2.255  log gf =     0.590 \\ 
 26    20    16    12     9     6     5    -1    -1    -1    -1    -1    -1    -1    -1    -1    -1    -1    -1    -1    -1    -1    -1    -1    -1    -1    -1 \\ 
 0.12  0.11  0.10  0.10  0.10  0.09  0.09 -1.00 -1.00 -1.00 -1.00 -1.00 -1.00 -1.00 -1.00 -1.00 -1.00 -1.00 -1.00 -1.00 -1.00 -1.00 -1.00 -1.00 -1.00 -1.00 -1.00 \\ 
 4913.6152  A  Ti1  Eexc =  1.872  log gf =     0.220 \\ 
 25    18    14    10     7     5    -1    -1    -1    -1    -1    -1    -1    -1    -1    -1    -1    -1    -1    -1    -1    -1    -1    -1    -1    -1    -1 \\ 
 0.11  0.10  0.10  0.10  0.09  0.09 -1.00 -1.00 -1.00 -1.00 -1.00 -1.00 -1.00 -1.00 -1.00 -1.00 -1.00 -1.00 -1.00 -1.00 -1.00 -1.00 -1.00 -1.00 -1.00 -1.00 -1.00 \\ 
 4981.7310  A  Ti1  Eexc =  0.848  log gf =     0.570 \\ 
 106    94    82    70    58    46    34    23    15     9     6    -1    -1    -1    -1    -1    -1    -1    -1    -1    -1    -1    -1    -1    -1    -1    -1 \\ 
 -0.02 -0.01 -0.00  0.02  0.03  0.05  0.07  0.08  0.09  0.09  0.09 -1.00 -1.00 -1.00 -1.00 -1.00 -1.00 -1.00 -1.00 -1.00 -1.00 -1.00 -1.00 -1.00 -1.00 -1.00 -1.00 \\ 
 4999.5020  A  Ti1  Eexc =  0.826  log gf =     0.320 \\ 
 89    77    65    53    42    31    22    14     9     6    -1    -1    -1    -1    -1    -1    -1    -1    -1    -1    -1    -1    -1    -1    -1    -1    -1 \\ 
 0.01  0.02  0.03  0.04  0.05  0.06  0.07  0.08  0.09  0.09 -1.00 -1.00 -1.00 -1.00 -1.00 -1.00 -1.00 -1.00 -1.00 -1.00 -1.00 -1.00 -1.00 -1.00 -1.00 -1.00 -1.00 \\ 
 5016.1611  A  Ti1  Eexc =  0.848  log gf =    -0.480 \\ 
 36    27    19    14     9     6    -1    -1    -1    -1    -1    -1    -1    -1    -1    -1    -1    -1    -1    -1    -1    -1    -1    -1    -1    -1    -1 \\ 
 0.09  0.08  0.07  0.06  0.06  0.07 -1.00 -1.00 -1.00 -1.00 -1.00 -1.00 -1.00 -1.00 -1.00 -1.00 -1.00 -1.00 -1.00 -1.00 -1.00 -1.00 -1.00 -1.00 -1.00 -1.00 -1.00 \\ 
 5025.5698  A  Ti1  Eexc =  2.041  log gf =     0.250 \\ 
 20    15    12     9     6     5    -1    -1    -1    -1    -1    -1    -1    -1    -1    -1    -1    -1    -1    -1    -1    -1    -1    -1    -1    -1    -1 \\ 
 0.16  0.14  0.12  0.11  0.09  0.08 -1.00 -1.00 -1.00 -1.00 -1.00 -1.00 -1.00 -1.00 -1.00 -1.00 -1.00 -1.00 -1.00 -1.00 -1.00 -1.00 -1.00 -1.00 -1.00 -1.00 -1.00 \\ 
 5036.4639  A  Ti1  Eexc =  1.443  log gf =     0.140 \\ 
 41    31    24    17    13     9     6    -1    -1    -1    -1    -1    -1    -1    -1    -1    -1    -1    -1    -1    -1    -1    -1    -1    -1    -1    -1 \\ 
 0.10  0.09  0.09  0.09  0.08  0.08  0.08 -1.00 -1.00 -1.00 -1.00 -1.00 -1.00 -1.00 -1.00 -1.00 -1.00 -1.00 -1.00 -1.00 -1.00 -1.00 -1.00 -1.00 -1.00 -1.00 -1.00 \\ 
 5173.7402  A  Ti1  Eexc =  0.000  log gf =    -1.060 \\ 
 41    29    20    13     9     6    -1    -1    -1    -1    -1    -1    -1    -1    -1    -1    -1    -1    -1    -1    -1    -1    -1    -1    -1    -1    -1 \\ 
 0.18  0.17  0.16  0.16  0.15  0.14 -1.00 -1.00 -1.00 -1.00 -1.00 -1.00 -1.00 -1.00 -1.00 -1.00 -1.00 -1.00 -1.00 -1.00 -1.00 -1.00 -1.00 -1.00 -1.00 -1.00 -1.00 \\ 
 5192.9692  A  Ti1  Eexc =  0.021  log gf =    -0.950 \\ 
 46    33    23    16    11     7     5    -1    -1    -1    -1    -1    -1    -1    -1    -1    -1    -1    -1    -1    -1    -1    -1    -1    -1    -1    -1 \\ 
 0.19  0.19  0.17  0.16  0.15  0.14  0.13 -1.00 -1.00 -1.00 -1.00 -1.00 -1.00 -1.00 -1.00 -1.00 -1.00 -1.00 -1.00 -1.00 -1.00 -1.00 -1.00 -1.00 -1.00 -1.00 -1.00 \\ 
 5210.3838  A  Ti1  Eexc =  0.048  log gf =    -0.820 \\ 
 54    40    29    20    14     9     6    -1    -1    -1    -1    -1    -1    -1    -1    -1    -1    -1    -1    -1    -1    -1    -1    -1    -1    -1    -1 \\ 
 0.17  0.17  0.17  0.16  0.15  0.14  0.14 -1.00 -1.00 -1.00 -1.00 -1.00 -1.00 -1.00 -1.00 -1.00 -1.00 -1.00 -1.00 -1.00 -1.00 -1.00 -1.00 -1.00 -1.00 -1.00 -1.00 \\ 
 5866.4492  A  Ti1  Eexc =  1.066  log gf =    -0.790 \\ 
 14    10     7     5    -1    -1    -1    -1    -1    -1    -1    -1    -1    -1    -1    -1    -1    -1    -1    -1    -1    -1    -1    -1    -1    -1    -1 \\ 
 0.15  0.14  0.13  0.13 -1.00 -1.00 -1.00 -1.00 -1.00 -1.00 -1.00 -1.00 -1.00 -1.00 -1.00 -1.00 -1.00 -1.00 -1.00 -1.00 -1.00 -1.00 -1.00 -1.00 -1.00 -1.00 -1.00 \\ 
 6258.0991  A  Ti1  Eexc =  1.443  log gf =    -0.390 \\ 
 19    14    10     7     5    -1    -1    -1    -1    -1    -1    -1    -1    -1    -1    -1    -1    -1    -1    -1    -1    -1    -1    -1    -1    -1    -1 \\ 
 0.08  0.06  0.05  0.04  0.04 -1.00 -1.00 -1.00 -1.00 -1.00 -1.00 -1.00 -1.00 -1.00 -1.00 -1.00 -1.00 -1.00 -1.00 -1.00 -1.00 -1.00 -1.00 -1.00 -1.00 -1.00 -1.00 \\ 
 6261.0962  A  Ti1  Eexc =  1.429  log gf =    -0.530 \\ 
 14    10     7     5    -1    -1    -1    -1    -1    -1    -1    -1    -1    -1    -1    -1    -1    -1    -1    -1    -1    -1    -1    -1    -1    -1    -1 \\ 
 0.08  0.06  0.05  0.04 -1.00 -1.00 -1.00 -1.00 -1.00 -1.00 -1.00 -1.00 -1.00 -1.00 -1.00 -1.00 -1.00 -1.00 -1.00 -1.00 -1.00 -1.00 -1.00 -1.00 -1.00 -1.00 -1.00 \\ 
 8426.5059  A  Ti1  Eexc =  0.826  log gf =    -1.200 \\ 
 16    11     7     5    -1    -1    -1    -1    -1    -1    -1    -1    -1    -1    -1    -1    -1    -1    -1    -1    -1    -1    -1    -1    -1    -1    -1 \\ 
 0.03  0.03  0.02  0.02 -1.00 -1.00 -1.00 -1.00 -1.00 -1.00 -1.00 -1.00 -1.00 -1.00 -1.00 -1.00 -1.00 -1.00 -1.00 -1.00 -1.00 -1.00 -1.00 -1.00 -1.00 -1.00 -1.00 \\ 
 2827.1140  A  Ti2  Eexc =  3.687  log gf =    -0.020 \\ 
 47    44    41    38    35    31    28    24    20    17    14    12    10     8     7     6     5    -1    -1    -1    -1    -1    -1    -1    -1    -1    -1 \\ 
 0.01  0.01  0.01  0.00  0.00 -0.00 -0.00 -0.00  0.00  0.00  0.01  0.01  0.02  0.03  0.04  0.05  0.07 -1.00 -1.00 -1.00 -1.00 -1.00 -1.00 -1.00 -1.00 -1.00 -1.00 \\ 
 2828.0769  A  Ti2  Eexc =  3.749  log gf =     0.870 \\ 
 80    78    75    72    69    66    62    58    54    49    45    40    37    33    30    27    24    22    19    16    14    12     9     8     6     5    -1 \\ 
 0.04  0.03  0.02  0.01 -0.01 -0.02 -0.03 -0.03 -0.03 -0.03 -0.02 -0.01  0.01  0.02  0.04  0.06  0.08  0.09  0.12  0.16  0.20  0.24  0.28  0.32  0.35  0.38 -1.00 \\ 
 2834.0110  A  Ti2  Eexc =  3.716  log gf =     0.000 \\ 
 48    45    42    39    36    32    29    25    21    18    15    12    10     8     7     6     5    -1    -1    -1    -1    -1    -1    -1    -1    -1    -1 \\ 
 0.01  0.01  0.01  0.00  0.00 -0.00 -0.00 -0.00 -0.00  0.00  0.01  0.01  0.02  0.03  0.04  0.05  0.07 -1.00 -1.00 -1.00 -1.00 -1.00 -1.00 -1.00 -1.00 -1.00 -1.00 \\ 
 2841.9351  A  Ti2  Eexc =  0.607  log gf =    -0.590 \\ 
 130   122   115   109   104    98    93    87    82    76    70    65    60    55    51    47    42    37    33    28    23    18    14    11     9     7     5 \\ 
 -0.01 -0.03 -0.04 -0.05 -0.07 -0.09 -0.10 -0.12 -0.13 -0.13 -0.14 -0.14 -0.13 -0.12 -0.10 -0.08 -0.06 -0.04 -0.01  0.02  0.07  0.11  0.16  0.20  0.24  0.28  0.30 \\ 
 2851.1011  A  Ti2  Eexc =  1.221  log gf =    -0.730 \\ 
 100    96    92    87    82    78    73    67    61    55    49    43    38    33    29    25    21    18    14    12     9     7     5    -1    -1    -1    -1 \\ 
 -0.03 -0.04 -0.05 -0.06 -0.08 -0.09 -0.09 -0.09 -0.09 -0.09 -0.08 -0.07 -0.05 -0.04 -0.03 -0.01  0.00  0.02  0.05  0.08  0.12  0.17  0.21 -1.00 -1.00 -1.00 -1.00 \\ 
 2853.9309  A  Ti2  Eexc =  0.607  log gf =    -1.550 \\ 
 91    87    82    77    71    66    60    54    47    40    33    28    23    19    15    12    10     8     6     5    -1    -1    -1    -1    -1    -1    -1 \\ 
 -0.02 -0.03 -0.04 -0.04 -0.05 -0.05 -0.05 -0.05 -0.04 -0.04 -0.04 -0.04 -0.03 -0.03 -0.03 -0.02 -0.01  0.00  0.03  0.05 -1.00 -1.00 -1.00 -1.00 -1.00 -1.00 -1.00 \\ 
 2868.7410  A  Ti2  Eexc =  0.574  log gf =    -1.380 \\ 
 97    93    88    83    78    72    66    60    54    47    41    35    29    25    20    17    14    11     9     7     5    -1    -1    -1    -1    -1    -1 \\ 
 -0.02 -0.03 -0.04 -0.05 -0.06 -0.06 -0.07 -0.07 -0.07 -0.06 -0.06 -0.06 -0.05 -0.05 -0.04 -0.03 -0.02 -0.01  0.01  0.04  0.08 -1.00 -1.00 -1.00 -1.00 -1.00 -1.00 \\ 
 4012.3850  A  Ti2  Eexc =  0.574  log gf =    -1.780 \\ 
 117   113   108   104    99    95    90    84    77    68    59    51    43    36    29    24    19    15    12     9     7     5    -1    -1    -1    -1    -1 \\ 
 -0.03 -0.03 -0.03 -0.03 -0.03 -0.02 -0.02 -0.02 -0.01 -0.01 -0.01 -0.01 -0.01 -0.01 -0.01 -0.00  0.01  0.02  0.04  0.07  0.10  0.13 -1.00 -1.00 -1.00 -1.00 -1.00 \\ 
 4028.3430  A  Ti2  Eexc =  1.891  log gf =    -0.920 \\ 
 105   103   100    97    94    90    87    82    76    69    62    54    47    41    35    30    25    21    17    14    11     8     6     5    -1    -1    -1 \\ 
 -0.03 -0.03 -0.03 -0.03 -0.03 -0.03 -0.03 -0.02 -0.02 -0.02 -0.01 -0.01 -0.01 -0.01 -0.00  0.00  0.01  0.02  0.04  0.06  0.09  0.12  0.15  0.17 -1.00 -1.00 -1.00 \\ 
 4053.8201  A  Ti2  Eexc =  1.892  log gf =    -1.070 \\ 
 98    95    92    89    86    82    78    73    67    60    52    45    38    32    27    23    19    16    12    10     8     6     5    -1    -1    -1    -1 \\ 
 -0.02 -0.03 -0.03 -0.03 -0.03 -0.02 -0.02 -0.01 -0.01 -0.01 -0.01 -0.01 -0.01 -0.01 -0.00  0.00  0.01  0.02  0.04  0.06  0.09  0.12  0.15 -1.00 -1.00 -1.00 -1.00 \\ 
 4161.5298  A  Ti2  Eexc =  1.084  log gf =    -2.090 \\ 
 84    80    75    70    65    59    52    45    38    31    24    19    15    12    10     7     6     5    -1    -1    -1    -1    -1    -1    -1    -1    -1 \\ 
 -0.02 -0.02 -0.02 -0.02 -0.01 -0.01 -0.01 -0.01 -0.01 -0.01 -0.01 -0.01 -0.00 -0.00  0.00  0.01  0.01  0.03 -1.00 -1.00 -1.00 -1.00 -1.00 -1.00 -1.00 -1.00 -1.00 \\ 
 4163.6401  A  Ti2  Eexc =  2.589  log gf =    -0.130 \\ 
 116   114   112   110   108   106   103    99    95    89    83    77    70    64    57    51    45    39    33    28    23    19    15    12    10     8     7 \\ 
 -0.04 -0.04 -0.05 -0.05 -0.06 -0.05 -0.05 -0.05 -0.04 -0.04 -0.03 -0.02 -0.01 -0.01  0.00  0.02  0.03  0.05  0.07  0.09  0.13  0.16  0.19  0.22  0.24  0.26  0.28 \\ 
 4174.0698  A  Ti2  Eexc =  2.598  log gf =    -1.260 \\ 
 54    53    50    47    44    40    37    32    28    23    19    15    13    10     8     7     6     5    -1    -1    -1    -1    -1    -1    -1    -1    -1 \\ 
 0.00 -0.00 -0.00 -0.00 -0.00 -0.00 -0.00 -0.00 -0.00 -0.00 -0.00  0.00  0.00  0.01  0.01  0.02  0.03  0.05 -1.00 -1.00 -1.00 -1.00 -1.00 -1.00 -1.00 -1.00 -1.00 \\ 
 4188.9868  A  Ti2  Eexc =  5.423  log gf =    -0.600 \\ 
 -1    -1    -1    -1    -1    -1    -1    -1    -1    -1    -1    -1    -1    -1    -1    -1    -1    -1    -1    -1    -1    -1    -1    -1    -1    -1    -1 \\ 
 -1.00 -1.00 -1.00 -1.00 -1.00 -1.00 -1.00 -1.00 -1.00 -1.00 -1.00 -1.00 -1.00 -1.00 -1.00 -1.00 -1.00 -1.00 -1.00 -1.00 -1.00 -1.00 -1.00 -1.00 -1.00 -1.00 -1.00 \\ 
 4190.2329  A  Ti2  Eexc =  1.084  log gf =    -3.122 \\ 
 24    21    18    15    12    10     8     7     5    -1    -1    -1    -1    -1    -1    -1    -1    -1    -1    -1    -1    -1    -1    -1    -1    -1    -1 \\ 
 -0.00 -0.00 -0.00 -0.00 -0.00 -0.00 -0.00 -0.00 -0.00 -1.00 -1.00 -1.00 -1.00 -1.00 -1.00 -1.00 -1.00 -1.00 -1.00 -1.00 -1.00 -1.00 -1.00 -1.00 -1.00 -1.00 -1.00 \\ 
 4287.8701  A  Ti2  Eexc =  1.080  log gf =    -1.790 \\ 
 103    99    95    90    85    80    74    67    59    50    41    34    27    22    18    14    11     9     7     5    -1    -1    -1    -1    -1    -1    -1 \\ 
 -0.07 -0.07 -0.06 -0.05 -0.05 -0.04 -0.03 -0.02 -0.02 -0.02 -0.01 -0.01 -0.01 -0.01 -0.00  0.00  0.01  0.03  0.05  0.07 -1.00 -1.00 -1.00 -1.00 -1.00 -1.00 -1.00 \\ 
 4290.2148  A  Ti2  Eexc =  1.164  log gf =    -0.870 \\ 
 148   143   138   133   129   125   120   115   109   102    95    87    80    73    65    58    50    43    36    30    24    19    14    11     9     7     5 \\ 
 -0.07 -0.08 -0.09 -0.09 -0.09 -0.08 -0.08 -0.07 -0.06 -0.06 -0.05 -0.05 -0.04 -0.03 -0.02 -0.01  0.00  0.01  0.04  0.06  0.10  0.13  0.17  0.20  0.23  0.26  0.27 \\ 
 4300.0488  A  Ti2  Eexc =  1.180  log gf =    -0.460 \\ 
 176   169   162   156   151   146   141   136   130   123   116   110   103    97    90    83    76    69    61    54    45    37    30    24    19    16    13 \\ 
 -0.06 -0.07 -0.08 -0.09 -0.10 -0.10 -0.10 -0.10 -0.10 -0.09 -0.09 -0.09 -0.08 -0.07 -0.06 -0.05 -0.03 -0.01  0.02  0.05  0.09  0.12  0.16  0.20  0.23  0.25  0.27 \\ 
 4301.9199  A  Ti2  Eexc =  1.160  log gf =    -1.210 \\ 
 129   125   120   116   112   107   102    97    91    83    75    66    58    51    43    37    30    25    20    16    12     9     7     5    -1    -1    -1 \\ 
 -0.07 -0.07 -0.07 -0.07 -0.06 -0.06 -0.05 -0.04 -0.04 -0.03 -0.03 -0.02 -0.02 -0.01 -0.01  0.00  0.01  0.02  0.05  0.07  0.10  0.14  0.17  0.20 -1.00 -1.00 -1.00 \\ 
 4374.8198  A  Ti2  Eexc =  2.060  log gf =    -1.570 \\ 
 65    62    59    55    51    47    42    37    31    25    21    17    13    11     9     7     6     5    -1    -1    -1    -1    -1    -1    -1    -1    -1 \\ 
 -0.00 -0.00 -0.01 -0.01 -0.01 -0.01 -0.01 -0.00 -0.00 -0.00 -0.01 -0.01 -0.00 -0.00 -0.00  0.00  0.01  0.02 -1.00 -1.00 -1.00 -1.00 -1.00 -1.00 -1.00 -1.00 -1.00 \\ 
 4386.8442  A  Ti2  Eexc =  2.598  log gf =    -0.960 \\ 
 75    73    71    68    65    61    57    52    46    39    33    28    23    19    16    13    11     9     7     6    -1    -1    -1    -1    -1    -1    -1 \\ 
 -0.01 -0.02 -0.02 -0.02 -0.02 -0.02 -0.01 -0.01 -0.01 -0.01 -0.00 -0.00  0.00  0.00  0.01  0.02  0.03  0.04  0.07  0.09 -1.00 -1.00 -1.00 -1.00 -1.00 -1.00 -1.00 \\ 
 4391.0200  A  Ti2  Eexc =  1.231  log gf =    -2.300 \\ 
 66    61    56    51    45    39    34    28    23    18    14    11     8     7     5    -1    -1    -1    -1    -1    -1    -1    -1    -1    -1    -1    -1 \\ 
 -0.01 -0.01 -0.01 -0.01 -0.01 -0.01 -0.01 -0.01 -0.00 -0.00 -0.00 -0.00 -0.00 -0.00 -0.00 -1.00 -1.00 -1.00 -1.00 -1.00 -1.00 -1.00 -1.00 -1.00 -1.00 -1.00 -1.00 \\ 
 4394.0591  A  Ti2  Eexc =  1.221  log gf =    -1.770 \\ 
 97    93    89    84    79    74    68    61    53    45    37    30    24    19    16    12    10     8     6     5    -1    -1    -1    -1    -1    -1    -1 \\ 
 -0.04 -0.04 -0.04 -0.03 -0.03 -0.02 -0.02 -0.01 -0.01 -0.01 -0.01 -0.01 -0.01 -0.00  0.00  0.01  0.02  0.03  0.05  0.08 -1.00 -1.00 -1.00 -1.00 -1.00 -1.00 -1.00 \\ 
 4395.0308  A  Ti2  Eexc =  1.084  log gf =    -0.540 \\ 
 176   169   162   156   151   146   142   137   131   124   117   110   103    97    90    83    76    68    60    52    43    36    28    23    18    15    12 \\ 
 -0.09 -0.10 -0.12 -0.13 -0.13 -0.13 -0.13 -0.12 -0.12 -0.12 -0.11 -0.11 -0.10 -0.09 -0.07 -0.06 -0.04 -0.02  0.01  0.05  0.09  0.12  0.16  0.20  0.22  0.25  0.26 \\ 
 4395.8389  A  Ti2  Eexc =  1.242  log gf =    -1.930 \\ 
 88    84    79    74    69    63    57    50    43    35    28    22    18    14    11     9     7     6    -1    -1    -1    -1    -1    -1    -1    -1    -1 \\ 
 -0.03 -0.03 -0.02 -0.02 -0.02 -0.01 -0.01 -0.01 -0.01 -0.01 -0.01 -0.01 -0.01 -0.01 -0.00  0.00  0.01  0.02 -1.00 -1.00 -1.00 -1.00 -1.00 -1.00 -1.00 -1.00 -1.00 \\ 
 4399.7720  A  Ti2  Eexc =  1.236  log gf =    -1.200 \\ 
 128   124   120   116   111   107   102    96    90    82    74    65    57    49    42    35    29    24    19    15    12     9     7     5    -1    -1    -1 \\ 
 -0.07 -0.08 -0.08 -0.07 -0.07 -0.06 -0.05 -0.05 -0.04 -0.03 -0.03 -0.02 -0.02 -0.01 -0.01  0.00  0.01  0.03  0.05  0.07  0.11  0.14  0.18  0.21 -1.00 -1.00 -1.00 \\ 
 4409.2349  A  Ti2  Eexc =  1.242  log gf =    -2.780 \\ 
 35    32    28    24    20    17    14    11     9     7     5    -1    -1    -1    -1    -1    -1    -1    -1    -1    -1    -1    -1    -1    -1    -1    -1 \\ 
 -0.00 -0.00 -0.00 -0.00 -0.00 -0.00 -0.00 -0.00 -0.00 -0.00 -0.00 -1.00 -1.00 -1.00 -1.00 -1.00 -1.00 -1.00 -1.00 -1.00 -1.00 -1.00 -1.00 -1.00 -1.00 -1.00 -1.00 \\ 
 4409.5200  A  Ti2  Eexc =  1.231  log gf =    -2.530 \\ 
 51    46    41    36    32    27    23    18    15    11     9     7     5    -1    -1    -1    -1    -1    -1    -1    -1    -1    -1    -1    -1    -1    -1 \\ 
 -0.01 -0.01 -0.01 -0.01 -0.01 -0.00 -0.00 -0.00 -0.00 -0.00 -0.00 -0.00 -0.00 -1.00 -1.00 -1.00 -1.00 -1.00 -1.00 -1.00 -1.00 -1.00 -1.00 -1.00 -1.00 -1.00 -1.00 \\ 
 4411.0698  A  Ti2  Eexc =  3.093  log gf =    -0.650 \\ 
 67    66    64    62    60    56    53    48    43    38    32    27    23    20    17    14    12    10     8     7     5    -1    -1    -1    -1    -1    -1 \\ 
 -0.01 -0.01 -0.01 -0.01 -0.01 -0.01 -0.01 -0.00 -0.00  0.00  0.00 -0.00  0.00  0.00  0.00  0.00  0.01  0.01  0.03  0.05  0.07 -1.00 -1.00 -1.00 -1.00 -1.00 -1.00 \\ 
 4411.9248  A  Ti2  Eexc =  1.224  log gf =    -2.620 \\ 
 45    41    36    31    27    23    19    15    12     9     7     5    -1    -1    -1    -1    -1    -1    -1    -1    -1    -1    -1    -1    -1    -1    -1 \\ 
 -0.01 -0.01 -0.01 -0.01 -0.00 -0.00 -0.00 -0.00 -0.00 -0.00 -0.00 -0.00 -1.00 -1.00 -1.00 -1.00 -1.00 -1.00 -1.00 -1.00 -1.00 -1.00 -1.00 -1.00 -1.00 -1.00 -1.00 \\ 
 4417.7129  A  Ti2  Eexc =  1.165  log gf =    -1.190 \\ 
 133   129   125   120   116   111   106   101    94    86    77    69    61    53    45    38    32    26    21    17    13    10     7     6    -1    -1    -1 \\ 
 -0.11 -0.12 -0.12 -0.11 -0.10 -0.09 -0.08 -0.07 -0.06 -0.05 -0.04 -0.03 -0.02 -0.02 -0.01 -0.00  0.01  0.02  0.05  0.07  0.10  0.14  0.17  0.20 -1.00 -1.00 -1.00 \\ 
 4418.3311  A  Ti2  Eexc =  1.236  log gf =    -1.990 \\ 
 85    81    76    71    66    60    53    46    39    32    25    20    16    13    10     8     6     5    -1    -1    -1    -1    -1    -1    -1    -1    -1 \\ 
 -0.03 -0.02 -0.02 -0.02 -0.02 -0.01 -0.01 -0.01 -0.01 -0.01 -0.01 -0.01 -0.00 -0.00  0.00  0.01  0.02  0.03 -1.00 -1.00 -1.00 -1.00 -1.00 -1.00 -1.00 -1.00 -1.00 \\ 
 4421.9380  A  Ti2  Eexc =  2.060  log gf =    -1.640 \\ 
 61    59    55    51    47    43    38    33    28    23    18    15    12     9     8     6     5    -1    -1    -1    -1    -1    -1    -1    -1    -1    -1 \\ 
 -0.01 -0.01 -0.01 -0.01 -0.01 -0.01 -0.01 -0.01 -0.01 -0.01 -0.01 -0.01 -0.01 -0.01 -0.00  0.00  0.01 -1.00 -1.00 -1.00 -1.00 -1.00 -1.00 -1.00 -1.00 -1.00 -1.00 \\ 
 4423.2388  A  Ti2  Eexc =  1.231  log gf =    -3.066 \\ 
 22    19    16    14    11     9     8     6     5    -1    -1    -1    -1    -1    -1    -1    -1    -1    -1    -1    -1    -1    -1    -1    -1    -1    -1 \\ 
 -0.00 -0.00 -0.00 -0.00 -0.00 -0.00 -0.00 -0.00 -0.00 -1.00 -1.00 -1.00 -1.00 -1.00 -1.00 -1.00 -1.00 -1.00 -1.00 -1.00 -1.00 -1.00 -1.00 -1.00 -1.00 -1.00 -1.00 \\ 
 4432.1089  A  Ti2  Eexc =  1.236  log gf =    -3.080 \\ 
 21    18    16    13    11     9     7     6     5    -1    -1    -1    -1    -1    -1    -1    -1    -1    -1    -1    -1    -1    -1    -1    -1    -1    -1 \\ 
 -0.00 -0.00 -0.00 -0.00 -0.00 -0.00 -0.00 -0.00 -0.00 -1.00 -1.00 -1.00 -1.00 -1.00 -1.00 -1.00 -1.00 -1.00 -1.00 -1.00 -1.00 -1.00 -1.00 -1.00 -1.00 -1.00 -1.00 \\ 
 4441.7300  A  Ti2  Eexc =  1.180  log gf =    -2.330 \\ 
 68    64    58    53    47    41    35    29    24    19    14    11     9     7     5    -1    -1    -1    -1    -1    -1    -1    -1    -1    -1    -1    -1 \\ 
 -0.03 -0.02 -0.02 -0.02 -0.02 -0.01 -0.01 -0.01 -0.01 -0.01 -0.01 -0.01 -0.01 -0.00  0.00 -1.00 -1.00 -1.00 -1.00 -1.00 -1.00 -1.00 -1.00 -1.00 -1.00 -1.00 -1.00 \\ 
 4443.8008  A  Ti2  Eexc =  1.080  log gf =    -0.710 \\ 
 164   158   152   147   142   138   133   128   122   115   108   101    94    87    80    72    65    57    49    41    34    27    21    17    13    10     8 \\ 
 -0.09 -0.11 -0.12 -0.12 -0.12 -0.12 -0.12 -0.11 -0.10 -0.10 -0.09 -0.09 -0.08 -0.06 -0.05 -0.04 -0.02 -0.00  0.03  0.06  0.09  0.13  0.17  0.20  0.23  0.25  0.26 \\ 
 4450.4819  A  Ti2  Eexc =  1.084  log gf =    -1.520 \\ 
 118   114   110   105   101    96    91    84    77    68    59    50    43    36    30    24    19    16    12     9     7     5    -1    -1    -1    -1    -1 \\ 
 -0.08 -0.08 -0.07 -0.07 -0.06 -0.05 -0.04 -0.04 -0.03 -0.02 -0.02 -0.02 -0.02 -0.01 -0.01 -0.00  0.01  0.02  0.05  0.07  0.10  0.14 -1.00 -1.00 -1.00 -1.00 -1.00 \\ 
 4464.4492  A  Ti2  Eexc =  1.161  log gf =    -1.810 \\ 
 100    96    91    86    81    75    69    62    54    45    37    30    24    19    16    12    10     8     6     5    -1    -1    -1    -1    -1    -1    -1 \\ 
 -0.06 -0.06 -0.06 -0.05 -0.04 -0.03 -0.03 -0.02 -0.02 -0.01 -0.01 -0.01 -0.01 -0.01 -0.00  0.00  0.01  0.03  0.05  0.08 -1.00 -1.00 -1.00 -1.00 -1.00 -1.00 -1.00 \\ 
 4468.5098  A  Ti2  Eexc =  1.130  log gf =    -0.630 \\ 
 197   187   177   169   162   154   148   141   133   125   117   109   101    94    87    79    71    64    55    47    39    31    25    20    16    12    10 \\ 
 -0.06 -0.07 -0.08 -0.09 -0.09 -0.09 -0.09 -0.09 -0.09 -0.09 -0.09 -0.09 -0.08 -0.07 -0.06 -0.04 -0.03 -0.01  0.02  0.05  0.09  0.13  0.17  0.20  0.23  0.25  0.27 \\ 
 4469.1509  A  Ti2  Eexc =  1.084  log gf =    -2.550 \\ 
 58    53    48    43    37    32    27    22    18    13    10     8     6     5    -1    -1    -1    -1    -1    -1    -1    -1    -1    -1    -1    -1    -1 \\ 
 -0.00 -0.00 -0.00 -0.00 -0.00 -0.00 -0.00 -0.00 -0.00 -0.00 -0.00 -0.01 -0.01 -0.00 -1.00 -1.00 -1.00 -1.00 -1.00 -1.00 -1.00 -1.00 -1.00 -1.00 -1.00 -1.00 -1.00 \\ 
 4470.8530  A  Ti2  Eexc =  1.165  log gf =    -2.020 \\ 
 88    83    79    73    68    61    55    48    40    33    26    21    16    13    10     8     6     5    -1    -1    -1    -1    -1    -1    -1    -1    -1 \\ 
 -0.05 -0.04 -0.04 -0.03 -0.03 -0.02 -0.02 -0.01 -0.01 -0.01 -0.01 -0.01 -0.01 -0.00 -0.00  0.01  0.02  0.03 -1.00 -1.00 -1.00 -1.00 -1.00 -1.00 -1.00 -1.00 -1.00 \\ 
 4488.3242  A  Ti2  Eexc =  3.122  log gf =    -0.500 \\ 
 76    75    73    71    69    66    63    58    53    47    41    35    30    26    22    19    16    14    11     9     7     6     5    -1    -1    -1    -1 \\ 
 -0.02 -0.02 -0.02 -0.02 -0.02 -0.01 -0.01 -0.01 -0.00 -0.00 -0.00 -0.00 -0.00 -0.00  0.00  0.00  0.01  0.01  0.03  0.05  0.07  0.09  0.12 -1.00 -1.00 -1.00 -1.00 \\ 
 4501.2700  A  Ti2  Eexc =  1.115  log gf =    -0.770 \\ 
 159   154   148   144   139   134   129   124   118   111   104    96    89    82    75    67    59    52    44    37    29    23    18    14    11     9     7 \\ 
 -0.10 -0.11 -0.12 -0.12 -0.12 -0.12 -0.11 -0.10 -0.10 -0.09 -0.08 -0.07 -0.06 -0.05 -0.04 -0.03 -0.01  0.00  0.03  0.06  0.10  0.13  0.17  0.20  0.23  0.25  0.27 \\ 
 4518.3301  A  Ti2  Eexc =  1.080  log gf =    -2.560 \\ 
 58    53    47    42    37    31    26    22    17    13    10     8     6     5    -1    -1    -1    -1    -1    -1    -1    -1    -1    -1    -1    -1    -1 \\ 
 -0.00 -0.00 -0.00 -0.00 -0.00 -0.00 -0.00 -0.00 -0.00 -0.00 -0.00 -0.01 -0.01 -0.00 -1.00 -1.00 -1.00 -1.00 -1.00 -1.00 -1.00 -1.00 -1.00 -1.00 -1.00 -1.00 -1.00 \\ 
 4529.4741  A  Ti2  Eexc =  1.571  log gf =    -1.750 \\ 
 83    79    75    70    65    60    54    47    40    33    27    21    17    14    11     9     7     6    -1    -1    -1    -1    -1    -1    -1    -1    -1 \\ 
 -0.04 -0.04 -0.03 -0.03 -0.02 -0.02 -0.02 -0.01 -0.01 -0.01 -0.01 -0.01 -0.01 -0.01 -0.01 -0.00  0.01  0.02 -1.00 -1.00 -1.00 -1.00 -1.00 -1.00 -1.00 -1.00 -1.00 \\ 
 4533.9600  A  Ti2  Eexc =  1.237  log gf =    -0.530 \\ 
 172   166   160   155   150   145   140   135   129   121   114   107   100    92    85    78    70    63    54    47    39    32    25    20    16    13    10 \\ 
 -0.12 -0.14 -0.15 -0.16 -0.16 -0.16 -0.15 -0.14 -0.14 -0.12 -0.12 -0.10 -0.09 -0.07 -0.05 -0.03 -0.01  0.01  0.04  0.06  0.10  0.14  0.18  0.21  0.24  0.26  0.28 \\ 
 4544.0200  A  Ti2  Eexc =  1.243  log gf =    -2.580 \\ 
 54    49    44    39    34    29    24    20    16    12     9     7     5    -1    -1    -1    -1    -1    -1    -1    -1    -1    -1    -1    -1    -1    -1 \\ 
 -0.00 -0.00 -0.00 -0.00 -0.00 -0.00 -0.00 -0.00 -0.00 -0.00 -0.00 -0.00 -0.00 -1.00 -1.00 -1.00 -1.00 -1.00 -1.00 -1.00 -1.00 -1.00 -1.00 -1.00 -1.00 -1.00 -1.00 \\ 
 4549.6201  A  Ti2  Eexc =  1.583  log gf =    -0.220 \\ 
 176   170   164   158   153   149   144   139   133   126   119   113   106   100    93    86    79    72    64    56    47    40    32    26    21    17    14 \\ 
 -0.11 -0.13 -0.14 -0.15 -0.15 -0.15 -0.15 -0.14 -0.14 -0.14 -0.13 -0.13 -0.12 -0.10 -0.09 -0.07 -0.05 -0.03 -0.00  0.03  0.08  0.12  0.16  0.20  0.23  0.26  0.28 \\ 
 4563.7568  A  Ti2  Eexc =  1.221  log gf =    -0.795 \\ 
 161   156   151   146   141   137   132   126   120   113   105    98    91    83    75    68    60    53    45    38    30    24    19    15    12     9     7 \\ 
 -0.12 -0.14 -0.15 -0.15 -0.15 -0.15 -0.14 -0.13 -0.12 -0.11 -0.10 -0.08 -0.07 -0.05 -0.04 -0.02 -0.00  0.01  0.04  0.07  0.11  0.14  0.18  0.21  0.24  0.26  0.28 \\ 
 4568.3140  A  Ti2  Eexc =  1.224  log gf =    -3.030 \\ 
 24    21    18    15    13    10     9     7     5    -1    -1    -1    -1    -1    -1    -1    -1    -1    -1    -1    -1    -1    -1    -1    -1    -1    -1 \\ 
 -0.01 -0.01 -0.01 -0.01 -0.01 -0.01 -0.01 -0.01 -0.01 -1.00 -1.00 -1.00 -1.00 -1.00 -1.00 -1.00 -1.00 -1.00 -1.00 -1.00 -1.00 -1.00 -1.00 -1.00 -1.00 -1.00 -1.00 \\ 
 4571.9712  A  Ti2  Eexc =  1.571  log gf =    -0.310 \\ 
 169   163   158   153   148   144   139   134   128   122   115   108   101    94    88    81    74    66    58    50    42    35    28    22    18    14    12 \\ 
 -0.12 -0.13 -0.14 -0.14 -0.15 -0.14 -0.14 -0.14 -0.13 -0.13 -0.12 -0.11 -0.10 -0.09 -0.08 -0.06 -0.04 -0.02  0.01  0.04  0.08  0.12  0.17  0.20  0.23  0.26  0.28 \\ 
 4583.4102  A  Ti2  Eexc =  1.164  log gf =    -2.840 \\ 
 36    32    28    24    20    17    14    11     9     7     5    -1    -1    -1    -1    -1    -1    -1    -1    -1    -1    -1    -1    -1    -1    -1    -1 \\ 
 -0.01 -0.01 -0.01 -0.01 -0.01 -0.01 -0.01 -0.00 -0.00 -0.00 -0.01 -1.00 -1.00 -1.00 -1.00 -1.00 -1.00 -1.00 -1.00 -1.00 -1.00 -1.00 -1.00 -1.00 -1.00 -1.00 -1.00 \\ 
 4589.9580  A  Ti2  Eexc =  1.237  log gf =    -1.620 \\ 
 109   105   101    96    91    85    80    73    65    56    47    39    32    26    21    17    14    11     8     6     5    -1    -1    -1    -1    -1    -1 \\ 
 -0.08 -0.08 -0.08 -0.07 -0.06 -0.05 -0.04 -0.03 -0.03 -0.02 -0.02 -0.01 -0.01 -0.01 -0.00  0.00  0.02  0.03  0.05  0.08  0.11 -1.00 -1.00 -1.00 -1.00 -1.00 -1.00 \\ 
 4636.3198  A  Ti2  Eexc =  1.165  log gf =    -3.024 \\ 
 26    23    20    17    14    12     9     8     6    -1    -1    -1    -1    -1    -1    -1    -1    -1    -1    -1    -1    -1    -1    -1    -1    -1    -1 \\ 
 0.00 -0.00 -0.00 -0.00 -0.00 -0.00 -0.00 -0.00 -0.00 -1.00 -1.00 -1.00 -1.00 -1.00 -1.00 -1.00 -1.00 -1.00 -1.00 -1.00 -1.00 -1.00 -1.00 -1.00 -1.00 -1.00 -1.00 \\ 
 4657.2012  A  Ti2  Eexc =  1.242  log gf =    -2.290 \\ 
 69    64    59    53    47    42    36    30    24    19    15    11     9     7     5    -1    -1    -1    -1    -1    -1    -1    -1    -1    -1    -1    -1 \\ 
 -0.02 -0.02 -0.02 -0.01 -0.01 -0.01 -0.01 -0.01 -0.01 -0.01 -0.01 -0.01 -0.01 -0.01 -0.01 -1.00 -1.00 -1.00 -1.00 -1.00 -1.00 -1.00 -1.00 -1.00 -1.00 -1.00 -1.00 \\ 
 4708.6631  A  Ti2  Eexc =  1.236  log gf =    -2.350 \\ 
 65    60    55    49    44    38    33    27    22    17    13    10     8     6     5    -1    -1    -1    -1    -1    -1    -1    -1    -1    -1    -1    -1 \\ 
 -0.02 -0.02 -0.02 -0.01 -0.01 -0.01 -0.01 -0.01 -0.01 -0.01 -0.01 -0.01 -0.01 -0.00 -0.00 -1.00 -1.00 -1.00 -1.00 -1.00 -1.00 -1.00 -1.00 -1.00 -1.00 -1.00 -1.00 \\ 
 4719.5151  A  Ti2  Eexc =  1.242  log gf =    -3.320 \\ 
 14    12    10     8     7     6     5    -1    -1    -1    -1    -1    -1    -1    -1    -1    -1    -1    -1    -1    -1    -1    -1    -1    -1    -1    -1 \\ 
 -0.01 -0.01 -0.01 -0.01 -0.01 -0.01 -0.01 -1.00 -1.00 -1.00 -1.00 -1.00 -1.00 -1.00 -1.00 -1.00 -1.00 -1.00 -1.00 -1.00 -1.00 -1.00 -1.00 -1.00 -1.00 -1.00 -1.00 \\ 
 4763.8799  A  Ti2  Eexc =  1.221  log gf =    -2.400 \\ 
 61    56    51    46    40    35    30    25    20    15    12     9     7     5    -1    -1    -1    -1    -1    -1    -1    -1    -1    -1    -1    -1    -1 \\ 
 -0.00 -0.00 -0.00 -0.00 -0.00 -0.00 -0.00 -0.00 -0.00 -0.00 -0.00 -0.00 -0.00 -0.00 -1.00 -1.00 -1.00 -1.00 -1.00 -1.00 -1.00 -1.00 -1.00 -1.00 -1.00 -1.00 -1.00 \\ 
 4764.5249  A  Ti2  Eexc =  1.236  log gf =    -2.690 \\ 
 42    38    34    29    25    21    17    14    11     8     6     5    -1    -1    -1    -1    -1    -1    -1    -1    -1    -1    -1    -1    -1    -1    -1 \\ 
 -0.00 -0.00 -0.00 -0.00 -0.00 -0.00 -0.00 -0.00 -0.00 -0.00 -0.00 -0.00 -1.00 -1.00 -1.00 -1.00 -1.00 -1.00 -1.00 -1.00 -1.00 -1.00 -1.00 -1.00 -1.00 -1.00 -1.00 \\ 
 4779.9849  A  Ti2  Eexc =  2.048  log gf =    -1.260 \\ 
 89    86    83    79    75    70    65    59    52    44    37    31    25    21    17    14    11     9     7     6    -1    -1    -1    -1    -1    -1    -1 \\ 
 -0.04 -0.05 -0.05 -0.04 -0.04 -0.03 -0.03 -0.02 -0.02 -0.02 -0.02 -0.01 -0.01 -0.01 -0.01 -0.00  0.00  0.01  0.03  0.05 -1.00 -1.00 -1.00 -1.00 -1.00 -1.00 -1.00 \\ 
 4798.5298  A  Ti2  Eexc =  1.080  log gf =    -2.660 \\ 
 53    48    43    37    32    27    23    18    15    11     8     6     5    -1    -1    -1    -1    -1    -1    -1    -1    -1    -1    -1    -1    -1    -1 \\ 
 -0.01 -0.01 -0.01 -0.01 -0.00 -0.00 -0.00 -0.00 -0.01 -0.01 -0.01 -0.01 -0.01 -1.00 -1.00 -1.00 -1.00 -1.00 -1.00 -1.00 -1.00 -1.00 -1.00 -1.00 -1.00 -1.00 -1.00 \\ 
 4805.0850  A  Ti2  Eexc =  2.061  log gf =    -0.960 \\ 
 107   104   101    98    94    90    85    79    73    64    56    48    42    35    30    25    21    17    14    11     8     7     5    -1    -1    -1    -1 \\ 
 -0.07 -0.07 -0.08 -0.07 -0.07 -0.06 -0.05 -0.04 -0.04 -0.03 -0.03 -0.02 -0.02 -0.02 -0.01 -0.01  0.00  0.01  0.03  0.05  0.08  0.11  0.15 -1.00 -1.00 -1.00 -1.00 \\ 
 4911.1899  A  Ti2  Eexc =  3.122  log gf =    -0.640 \\ 
 68    67    66    64    61    58    55    50    45    39    33    28    24    21    17    15    12    10     9     7     6     5    -1    -1    -1    -1    -1 \\ 
 0.01  0.01  0.00  0.00 -0.00 -0.00 -0.00 -0.00 -0.00 -0.00 -0.00 -0.00 -0.01 -0.01 -0.01 -0.01 -0.00  0.00  0.01  0.03  0.05  0.08 -1.00 -1.00 -1.00 -1.00 -1.00 \\ 
 4996.3672  A  Ti2  Eexc =  1.582  log gf =    -3.290 \\ 
 8     7     6     5    -1    -1    -1    -1    -1    -1    -1    -1    -1    -1    -1    -1    -1    -1    -1    -1    -1    -1    -1    -1    -1    -1    -1 \\ 
 -0.00 -0.00 -0.01 -0.01 -1.00 -1.00 -1.00 -1.00 -1.00 -1.00 -1.00 -1.00 -1.00 -1.00 -1.00 -1.00 -1.00 -1.00 -1.00 -1.00 -1.00 -1.00 -1.00 -1.00 -1.00 -1.00 -1.00 \\ 
 5005.1572  A  Ti2  Eexc =  1.565  log gf =    -2.730 \\ 
 25    23    20    17    15    12    10     8     6     5    -1    -1    -1    -1    -1    -1    -1    -1    -1    -1    -1    -1    -1    -1    -1    -1    -1 \\ 
 -0.01 -0.01 -0.01 -0.01 -0.01 -0.01 -0.01 -0.01 -0.01 -0.01 -1.00 -1.00 -1.00 -1.00 -1.00 -1.00 -1.00 -1.00 -1.00 -1.00 -1.00 -1.00 -1.00 -1.00 -1.00 -1.00 -1.00 \\ 
 5010.2100  A  Ti2  Eexc =  3.093  log gf =    -1.350 \\ 
 28    27    26    24    22    20    18    16    13    11     9     7     6     5    -1    -1    -1    -1    -1    -1    -1    -1    -1    -1    -1    -1    -1 \\ 
 -0.00 -0.00 -0.00 -0.00 -0.00 -0.00 -0.00 -0.00 -0.00 -0.00 -0.01 -0.01 -0.01 -0.01 -1.00 -1.00 -1.00 -1.00 -1.00 -1.00 -1.00 -1.00 -1.00 -1.00 -1.00 -1.00 -1.00 \\ 
 5013.3301  A  Ti2  Eexc =  3.095  log gf =    -2.028 \\ 
 7     7     7     6     6     5    -1    -1    -1    -1    -1    -1    -1    -1    -1    -1    -1    -1    -1    -1    -1    -1    -1    -1    -1    -1    -1 \\ 
 -0.00 -0.00 -0.00 -0.00 -0.00 -0.00 -1.00 -1.00 -1.00 -1.00 -1.00 -1.00 -1.00 -1.00 -1.00 -1.00 -1.00 -1.00 -1.00 -1.00 -1.00 -1.00 -1.00 -1.00 -1.00 -1.00 -1.00 \\ 
 5013.6860  A  Ti2  Eexc =  1.581  log gf =    -2.140 \\ 
 61    56    52    47    42    37    32    26    22    17    13    10     8     6     5    -1    -1    -1    -1    -1    -1    -1    -1    -1    -1    -1    -1 \\ 
 -0.02 -0.01 -0.01 -0.01 -0.01 -0.01 -0.01 -0.01 -0.01 -0.01 -0.01 -0.01 -0.01 -0.01 -0.00 -1.00 -1.00 -1.00 -1.00 -1.00 -1.00 -1.00 -1.00 -1.00 -1.00 -1.00 -1.00 \\ 
 5072.2900  A  Ti2  Eexc =  3.122  log gf =    -1.020 \\ 
 46    45    43    41    39    36    33    29    25    21    17    14    12    10     8     7     6     5    -1    -1    -1    -1    -1    -1    -1    -1    -1 \\ 
 -0.01 -0.01 -0.01 -0.01 -0.01 -0.01 -0.01 -0.01 -0.01 -0.01 -0.01 -0.01 -0.01 -0.01 -0.01 -0.01 -0.00  0.00 -1.00 -1.00 -1.00 -1.00 -1.00 -1.00 -1.00 -1.00 -1.00 \\ 
 5129.1602  A  Ti2  Eexc =  1.891  log gf =    -1.340 \\ 
 97    93    89    85    80    75    70    63    55    47    39    33    27    22    18    15    12    10     8     6     5    -1    -1    -1    -1    -1    -1 \\ 
 -0.07 -0.06 -0.06 -0.05 -0.05 -0.04 -0.03 -0.03 -0.03 -0.02 -0.02 -0.02 -0.02 -0.02 -0.02 -0.01 -0.00  0.00  0.02  0.05  0.07 -1.00 -1.00 -1.00 -1.00 -1.00 -1.00 \\ 
 5154.0698  A  Ti2  Eexc =  1.566  log gf =    -1.750 \\ 
 90    86    81    76    70    64    58    50    43    35    28    23    18    14    11     9     7     6    -1    -1    -1    -1    -1    -1    -1    -1    -1 \\ 
 -0.06 -0.06 -0.06 -0.05 -0.05 -0.04 -0.03 -0.03 -0.02 -0.02 -0.02 -0.02 -0.02 -0.01 -0.01 -0.00  0.01  0.02 -1.00 -1.00 -1.00 -1.00 -1.00 -1.00 -1.00 -1.00 -1.00 \\ 
 5185.9131  A  Ti2  Eexc =  1.892  log gf =    -1.410 \\ 
 92    89    85    80    76    70    65    58    51    43    35    29    24    19    16    13    10     8     7     5    -1    -1    -1    -1    -1    -1    -1 \\ 
 -0.06 -0.06 -0.05 -0.05 -0.04 -0.03 -0.03 -0.03 -0.02 -0.02 -0.02 -0.02 -0.02 -0.02 -0.01 -0.01 -0.00  0.01  0.02  0.05 -1.00 -1.00 -1.00 -1.00 -1.00 -1.00 -1.00 \\ 
 5188.6802  A  Ti2  Eexc =  1.582  log gf =    -1.050 \\ 
 134   130   126   121   117   111   106    99    92    83    74    64    56    48    41    34    28    23    19    15    11     9     7     5    -1    -1    -1 \\ 
 -0.15 -0.15 -0.15 -0.14 -0.14 -0.12 -0.11 -0.09 -0.08 -0.07 -0.06 -0.05 -0.04 -0.03 -0.02 -0.01  0.00  0.02  0.04  0.07  0.10  0.14  0.17  0.21 -1.00 -1.00 -1.00 \\ 
 5211.5361  A  Ti2  Eexc =  2.589  log gf =    -1.410 \\ 
 50    48    46    43    39    36    32    27    23    19    15    12    10     8     6     5    -1    -1    -1    -1    -1    -1    -1    -1    -1    -1    -1 \\ 
 -0.00 -0.00 -0.00 -0.01 -0.01 -0.01 -0.01 -0.00 -0.00 -0.00 -0.00 -0.00 -0.00  0.00  0.01  0.01 -1.00 -1.00 -1.00 -1.00 -1.00 -1.00 -1.00 -1.00 -1.00 -1.00 -1.00 \\ 
 5262.1401  A  Ti2  Eexc =  1.582  log gf =    -2.250 \\ 
 55    51    46    41    37    32    27    22    18    14    11     8     6     5    -1    -1    -1    -1    -1    -1    -1    -1    -1    -1    -1    -1    -1 \\ 
 -0.03 -0.03 -0.03 -0.02 -0.02 -0.02 -0.02 -0.02 -0.01 -0.01 -0.01 -0.01 -0.01 -0.01 -1.00 -1.00 -1.00 -1.00 -1.00 -1.00 -1.00 -1.00 -1.00 -1.00 -1.00 -1.00 -1.00 \\ 
 5268.6099  A  Ti2  Eexc =  2.597  log gf =    -1.610 \\ 
 38    36    34    31    28    25    22    19    16    13    10     8     6     5    -1    -1    -1    -1    -1    -1    -1    -1    -1    -1    -1    -1    -1 \\ 
 0.00 -0.00 -0.00 -0.00 -0.00 -0.00 -0.00 -0.00 -0.00 -0.00 -0.00 -0.00 -0.00  0.00 -1.00 -1.00 -1.00 -1.00 -1.00 -1.00 -1.00 -1.00 -1.00 -1.00 -1.00 -1.00 -1.00 \\ 
 5336.7861  A  Ti2  Eexc =  1.581  log gf =    -1.600 \\ 
 100    96    91    86    81    75    68    61    53    44    37    30    24    19    16    13    10     8     6     5    -1    -1    -1    -1    -1    -1    -1 \\ 
 -0.07 -0.07 -0.06 -0.06 -0.05 -0.04 -0.04 -0.03 -0.03 -0.02 -0.02 -0.02 -0.02 -0.02 -0.01 -0.01  0.00  0.02  0.04  0.06 -1.00 -1.00 -1.00 -1.00 -1.00 -1.00 -1.00 \\ 
 5381.0220  A  Ti2  Eexc =  1.565  log gf =    -1.970 \\ 
 75    71    66    60    55    49    43    36    30    24    19    15    12     9     7     6    -1    -1    -1    -1    -1    -1    -1    -1    -1    -1    -1 \\ 
 -0.04 -0.04 -0.03 -0.03 -0.03 -0.02 -0.02 -0.02 -0.02 -0.01 -0.01 -0.02 -0.01 -0.01 -0.01 -0.00 -1.00 -1.00 -1.00 -1.00 -1.00 -1.00 -1.00 -1.00 -1.00 -1.00 -1.00 \\ 
 5418.7681  A  Ti2  Eexc =  1.581  log gf =    -2.130 \\ 
 64    59    55    49    44    39    33    28    23    18    14    11     8     7     5    -1    -1    -1    -1    -1    -1    -1    -1    -1    -1    -1    -1 \\ 
 -0.03 -0.03 -0.02 -0.02 -0.02 -0.02 -0.02 -0.01 -0.01 -0.01 -0.01 -0.01 -0.01 -0.01 -0.01 -1.00 -1.00 -1.00 -1.00 -1.00 -1.00 -1.00 -1.00 -1.00 -1.00 -1.00 -1.00 \\ 
 6680.1328  A  Ti2  Eexc =  3.093  log gf =    -1.890 \\ 
 11    11    10     9     8     7     6     5    -1    -1    -1    -1    -1    -1    -1    -1    -1    -1    -1    -1    -1    -1    -1    -1    -1    -1    -1 \\ 
 -0.00 -0.01 -0.01 -0.01 -0.01 -0.01 -0.01 -0.01 -1.00 -1.00 -1.00 -1.00 -1.00 -1.00 -1.00 -1.00 -1.00 -1.00 -1.00 -1.00 -1.00 -1.00 -1.00 -1.00 -1.00 -1.00 -1.00 \\ 
 6998.9048  A  Ti2  Eexc =  3.122  log gf =    -1.280 \\ 
 36    34    33    30    28    25    22    19    16    13    10     8     7     6     5    -1    -1    -1    -1    -1    -1    -1    -1    -1    -1    -1    -1 \\ 
 -0.01 -0.01 -0.01 -0.01 -0.01 -0.02 -0.02 -0.02 -0.02 -0.03 -0.03 -0.04 -0.05 -0.05 -0.06 -1.00 -1.00 -1.00 -1.00 -1.00 -1.00 -1.00 -1.00 -1.00 -1.00 -1.00 -1.00 \\

\section*{Acknowledgments}

We thank Keith~Butler for computations of the photoionisation cross-sections for Ti\ii.
This research was supported by the Russian Foundation for Basic Research (grants 16-32-00695 and 15-02-06046).
TS and LM are grateful to the Swiss National Science Foundation (the SCOPES project IZ73Z0--152485).
TS and LM are indebted to the International Space Science Institute (ISSI), Bern, Switzerland, for supporting and funding the international team "First stars in dwarf galaxies" and "The Formation and Evolution of the Galactic Halo".
We made use of the NIST, SIMBAD, and VALD databases.


\bibliographystyle{mn2e}



\end{document}